\documentclass[10pt,doublecolumn,twoside]{IEEEtran}

\usepackage{kbordermatrix}

\renewcommand{\kbldelim}{(}
\renewcommand{\kbrdelim}{)}

\usepackage{etoolbox}
\newtoggle{doublecolumn}
\newtoggle{calculateFigures}

\toggletrue{doublecolumn}
\togglefalse{calculateFigures}
\usepackage{etex}

\usepackage[english]{babel}
\usepackage{lipsum}
\usepackage{amsbsy,amsmath,amsfonts,amssymb,amsthm}
\usepackage{mathtools}
\usepackage{textcomp} %degrees centigrade symbols, euros, etc. 
\usepackage{relsize}

\usepackage{bm,cite,cases,pstricks,times,url,verbatim} % try to remove this line for http://www.conf-express.org
\usepackage[noend]{algpseudocode}
\usepackage{booktabs}
\usepackage{tabularx,array,dcolumn,multirow}

\usepackage{yfonts}

\usepackage{dutchcal} % lower case calligraphic, boondox-cal

\usepackage{soul} %\ul per underline

\usepackage{blkarray,bigdelim} % matrices with braces

\allowdisplaybreaks[4]

\usepackage{centernot} % prints the sym­bol \not on the fol­low­ing ar­gu­ment

\newcommand{\subparagraph}{}

\iftoggle{doublecolumn}{
    \newtheorem{thm}{Theorem}
    \newtheorem{fact}{Fact}
    \newtheorem{lemma}{Lemma}
    \newtheorem{definition}{Definition}
    \newtheorem{conj}{Conjecture}
    \newtheorem{propos}{Proposition}
    \newtheorem{corol}{Corollary}
    \newtheorem{ass}{Assumption}
    \newtheorem{example}{Example}
    \newtheorem{remark}{Remark}
    \newtheorem{note}{Note}
    \newtheorem{obs}{Observation}
}{

    \newtheoremstyle{exampstyle}
      {0} % Space above
      {0} % Space below
      {\itshape} % Body font
      {} % Indent amount
      {\bfseries} % Theorem head font
      {.} % Punctuation after theorem head
      {.5em} % Space after theorem head
      {} % Theorem head spec (can be left empty, meaning `normal')

    \theoremstyle{exampstyle} \newtheorem{thm}{Theorem}
    \theoremstyle{exampstyle} 
    \theoremstyle{exampstyle} 
    \theoremstyle{exampstyle} 
    \theoremstyle{exampstyle} 
    \theoremstyle{exampstyle} \newtheorem{propos}{Proposition}
    \theoremstyle{exampstyle} \newtheorem{corol}{Corollary}
    \theoremstyle{exampstyle} 
    \theoremstyle{exampstyle} 
    \theoremstyle{exampstyle} 
    \theoremstyle{exampstyle} 
    \theoremstyle{exampstyle} 
}

\newcommand{\argmin}[1]{\underset{#1}{\operatorname{arg}\,\operatorname{min}}\;}

\makeatletter
\newcommand{\pushright}[1]{\ifmeasuring@#1\else\omit\hfill$\displaystyle#1$\fi\ignorespaces}
\newcommand{\pushleft}[1]{\ifmeasuring@#1\else\omit$\displaystyle#1$\hfill\fi\ignorespaces}

\begingroup
\catcode`\#=11
\gdef\noautorotate{-dAutoRotatePages#/None}
\endgroup

\usepackage{graphicx,xcolor,float,dblfloatfix}
\usepackage{psfrag}

\usepackage{caption}
\usepackage{subcaption}

\newcommand{\subalign}[1]{%
  \vcenter{%
    \Let@ \restore@math@cr \default@tag
    \baselineskip\fontdimen10 \scriptfont\tw@
    \advance\baselineskip\fontdimen12 \scriptfont\tw@
    \lineskip\thr@@\fontdimen8 \scriptfont\thr@@
    \lineskiplimit\lineskip
    \ialign{\hfil$\m@th\textstyle##$&$\m@th\textstyle{}##$\crcr
      #1\crcr
    }%
  }
}

\iftoggle{calculateFigures}{
    \usepackage[crop=off,runs=2,pspdf=\noautorotate]{auto-pst-pdf}
}{
    \usepackage[off]{auto-pst-pdf}
}

\usepackage{hyperref}

\graphicspath{%
{./Figures/}
}
\usepackage{caption}
\usepackage{subcaption}
\begin{document}

\author{\IEEEauthorblockN{Alessandro~Biason,~\IEEEmembership{Student~Member,~IEEE,} Urbashi~Mitra,~\IEEEmembership{Fellow,~IEEE,} and~Michele~Zorzi,~\IEEEmembership{Fellow,~IEEE}
\thanks{Alessandro~Biason and Michele~Zorzi are with the Department of Information Engineering, University of Padova - via Gradenigo
6b, 35131 Padova, Italy (email: biasonal@dei.unipd.it, zorzi@dei.unipd.it).}
\thanks{Urbashi~Mitra is with the Ming Hsieh Department of Electrical Engineering, University of Southern California, Los Angeles, USA (email: ubli@usc.edu).}
\thanks{A preliminary version of this paper has been accepted for presentation at IEEE ISIT 2016~\cite{Biason2016c}.}
}}

%\title{On the Performance of Active Sensing \\ with a Communication Channel}
\title{Improved Active Sensing Performance in Wireless Sensor Networks via Channel State Information - Extended Version}

\maketitle

\fontdimen2\font=3.5pt

\begin{abstract}
Active sensing refers to the process of choosing or tuning a set of sensors in order to track an underlying system in an efficient and accurate way. In a wireless environment, among the several kinds of features extracted by traditional sensors, the information carried by the communication channel about the state of the system can be used to further boost the tracking performance and save energy. A joint tracking problem which considers sensor measurements and communication channel together for tracking purposes is set up and solved. The system is modeled as a partially observable Markov decision problem and the properties of the cost-to-go function are used to reduce the problem complexity. In particular, upper and lower bounds to the optimal sensor selection choice are derived and used to introduce sub-optimal sensing strategies. Numerical results show the advantages of using the channel as an additional way for improving the tracking precision and reduce the energy costs.
\end{abstract}

\begin{IEEEkeywords}
communication channel, features, measurements, wireless body area network, wireless sensor networks, partially observable Markov decision problems, optimization, policies.
\end{IEEEkeywords}

\section{Introduction}

\IEEEPARstart{T}{racking} is a common application in Wireless Sensor Networks (WSNs) and in the Internet of Things (IoT) world in which different devices collaborate to detect a common underlying state of a system. Indeed, due to the dense nature of these networks and the limited computational capabilities and energy availability of the devices, redundant data from multiple sensors can be combined to improve the tracking accuracy.
Among all the physical quantities that can be exploited for tracking, the use of \emph{Channel State Information} (CSI) as a way to improve the detection performance has been studied only marginally to date and only in certain contexts (e.g., target localization~\cite{Paul2009}). However, since in many applications the channel is influenced by the underlying system, it can be exploited in a tracking system to improve its performance and reduce the energy costs. The goal of this paper is to investigate a \emph{joint}  active sensing optimization problem in which, in addition to the standard sensor measurements, the communication channel is also exploited.

Examples of active sensing applications are compressive spectrum sensing~\cite{Sun2013}, object tracking~\cite{Fuemmeler2011}, health care~\cite{Zhou2008} and sparse signal recovery~\cite{Wei2015}. Moreover, active sensing has been used in Wireless Body Area Networks (WBANs)~\cite{Thatte2011,Ghasemzadeh2013,Zois2013,Quwaider2008,Archasantisuk2015}, which we will also use as a practical scenario for our model. 
In a WBAN, different in-body and on-body sensors obtain noisy measurements of a quantity (e.g., features of the electrocardiogram or data from a multi-axes accelerometer) related to the current underlying and unknown physical activity of a subject (e.g., sitting, standing, walking, running, etc.), and collaborate by transmitting the gathered data to a common Fusion Center (FC) (e.g., a mobile phone in a pocket). The role of the FC is to combine the measurements and assess the current activity. This may be particularly useful for health-care applications (e.g., for epidemiologic/clinical research purposes~\cite{Zhou2008}) or for monitoring the daily physical activity of a patient.

Differently from other tracking activities (e.g., RADARS) wherein reliable, accurate and expensive, in a WSN sensors are employed and the tracking task is not trivial because of energy and/or computational constraints and inaccuracies in the measurements. Indeed, since the devices are generally battery powered, energy efficiency is a key aspect to consider, and prolonging the network lifetime represents one of the most studied challenges (see~\cite{Dietrich2009}). Among the traditional approaches to maximize the battery life, prediction-based schemes with active activation have received recent attention. The basic idea is to dynamically activate and use only the most useful subset of nodes in the network and turn off the others to save energy. The active nodes transmit their data to the FC, which estimates the underlying activity of the system and decides the portion of the network to activate in the next time slot. This question is known as the \emph{Sensor Selection Problem} (SSP)~\cite{Chen2007,Chepuri2015} and aims at minimizing the energy consumption for the sensors and the fusion center, while still providing a reliable estimate.

While obtaining accurate estimations is generally possible by increasing the number of gathered measurements, it also incurs high energy costs both at the sensor~\cite{Krishnamurthy2002,Krishnamurthy2007,Krishnamurthy2013,Au2012} and at the FC sides~\cite{Thatte2011,Zois2013}. In the KNOW-ME network considered in~\cite{Thatte2011}, the FC represents the performance bottleneck bacause of the data reception costs. The model in~\cite{Thatte2011} was extended in~\cite{Zois2013} to the case of an underlying dynamic process described by a Markov Chain and a POMDP framework was employed to solve the problem. Our paper extends~\cite{Zois2013} to the case in which a communication channel is considered. The POMDP approach had been previously considered by Krishnamurthy~\cite{Krishnamurthy2002,Krishnamurthy2007} but, differently from our approach, without accounting for the possibility of choosing multiple sensors simultaneously and using discrete measurements, or focusing on the problem of optimizing the number of samples to gather and not on the sensor selection~\cite{Krishnamurthy2013}. Reference~\cite{Au2012} set up a POMDP model for selecting sensor resources for context classification under an average energy consumption constraint but considering only the two control actions ``activate'' or ``deactivate'' all sensors. The authors of~\cite{Savage2009} studied a constrained sensor scheduling problem in the Gauss-Markov framework, and explicitly derived the optimal strategy. In~\cite{Savage2009}, the underlying system evolves in a Gaussian-fashion and is not subject to a Markov evolution. Also, cooperation between sensors is not taken into account.

However, most of the previous works did not explicitly consider the communication channel between sensors and fusion center~\cite{Quevedo2013}, neither it was used to improve the tracking performance. Using the channel as an additional feature results in several advantages. First, the channel information is intrinsically related with the reception of the sensor measurements, thus no additional energy costs are required to obtain it. Second, there may be situations in which the traditional measurements are less informative than the communication channel. Finally, when a sensor measurement is lost or highly corrupted by noise, because of a bad channel condition, it is still possible to gather information about the underlying state of the system (e.g., if for certain activities it is likely that a bad channel is observed, whereas others experience good channel conditions almost always, then a packet loss may be very informative about the underlying activity).

The communication channel was explicitly taken into account in~\cite{Gupta2006}, where Gupta \emph{et al.} showed how the packet drop probability influences the optimal sensor selection policy. In \cite{Wu2013}, the authors studied the relation between estimation quality and communication rate and showed that it is possible to significantly reduce the communication rate for a small degradation of the sensing performance. However, neither~\cite{Gupta2006} or~\cite{Wu2013} considered a POMDP framework and optimized the number of measurements to gather from every sensor. Limited bandwidth constraints of the channel were considered in~\cite{Shi2012}, in which an optimal scheduler was built for two independent Gauss-Markov systems. Similarly, \cite{Yang2013}~studied bandwidth constraints in a multi-dimensional system for both the homogeneous and heterogeneous cases.
The possibility to explicitly exploit the Received Signal Strength Indicator (RSSI) for body activity tracking purposes in a real scenario was described in~\cite{Quwaider2008}. Recently, \cite{Archasantisuk2015}~introduced a machine learning technique to achieve high detection accuracy using the RSSI. References~\cite{Quwaider2008} and~\cite{Archasantisuk2015} did not focus on the sensor selection optimization problem and use \emph{only} CSI for state detection. Instead, in this work, we enclose in a POMDP framework the communication channel and the traditional measurements of the sensors jointly. This makes the optimization more challenging to solve, thus addressing the problem requires new techniques and results.

\begin{figure}[!t]
  \centering
  \includegraphics[trim = 0mm 0mm 0mm 0mm, width=.8\columnwidth]{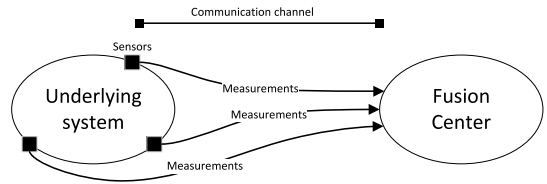}
  \caption{Block diagram of the system.}
  \label{fig:model}
\end{figure}

The main contributions of the paper can be summarized as follows. We set up an \emph{active sensing} model (see \figurename~\ref{fig:model}) in which, at every time step, a set of sensors is chosen to track the underlying state of the system. The optimal performance in terms of tracking quality and energy consumption is derived exploiting a POMDP framework for the infinite horizon setup, so that only a stationary scheduler needs to be stored in the nodes.
We assume that the sensors are \emph{passive} (i.e., they do not influence the underlying state) and \emph{heterogeneous} in terms of sensing cost, quality of the measurements and communication channel. As in~\cite{Thatte2011,Zois2013}, and differently from many previous works~\cite{Krishnamurthy2002,Krishnamurthy2007,Krishnamurthy2013,Au2012} we consider an energy constrained fusion center.
Since using the channel as an additional source of information adds a layer of complexity to the problem, we simplify the problem in different steps. First, with the goal of reducing the size of the belief space~\cite{Shani2013}, we use the concavity properties of the cost-to-go function, $J$, to derive a lower bound to $J$ (Corollary~\ref{corol:J_concave})~\cite{Smallwood1973}. This, in conjunction with an upper bound based on the tangents of $J$ can be used to (1) estimate the cost of the optimal sensing strategy, (2) introduce sub-optimal probabilistic tracking strategies and (3) classify these sub-optimal solutions. Then, we decompose the tracking procedure into a simpler set of operations (Theorem~\ref{thm:belief_simplified}) and cast a multi-dimensional problem in simpler uni-dimensional sub-problems. Finally, we propose a sub-optimal greedy technique which further simplifies the optimization process. 
Numerical results support the importance of considering channel and measurements jointly and validate our theoretical results. While we use the WBAN case as baseline for the numerical evaluations, the proposed model adopts very general assumptions and the theoretical framework can be applied to a large variety of applications (e.g., object tracking, indoor environmental monitoring, etc.).

The paper is organized as follows. Section~\ref{sec:system_model} defines the system model we analyze and introduces the notation. In Sections~\ref{sec:tracking} and~\ref{sec:optimization} we describe the tracking procedure and optimization problem. Section~\ref{sec:analysis} presents the main result of the paper in terms of bounds and sub-optimal policies. Section~\ref{sec:numerical_results} shows our numerical results. Finally, Section~\ref{sec:conclusions} concludes the paper.

%: advantages long-term. ``The obvious advantage of a stationary scheduler is that only the piecewise linear representation of $p$ and its associated optimal decision need be stored in memory for the real-time implementation.'' Faster to compute

\section{System Model} \label{sec:system_model}

\begin{table*}[t]
    \centering
    \caption{Notation and parameters.}
    \label{tab:parameters}
    \begin{tabular}{ c|l|l }
        \toprule
        & \textbf{Symbol} & \textbf{Meaning} \\
        \midrule
        \multirow{3}{*}{Notation} & $\mathrm{a}_{u,s,k},\mathrm{A}_{s}$ & scalar referred to the $u$-th measurement of sensor $s$ in slot $k$ and corresponding random variable\\
        & $\mathbf{a}_k, \mathbf{A}$ & set of all the scalars $\mathrm{a}_{u,s,k}$ in slot $k$ and corresponding random variable \\
        & $\mathcal{A}_k = \{\mathbf{a}_0,\ldots,\mathbf{a}_k\}$ & set of all the scalars $\mathrm{a}_{u,s,k}$ until slot $k$ \\
        \hline
        \multirow{4}{*}{Indices} & $s,k$ & sensor and slot indices \\
        & $u = 1,\ldots,N_s^{\mathbf{u}_{k-1}}$ & measurement index of a single sensor in a single slot \\
        & $\nu = 1,\ldots,\sum_s N_s^{\mathbf{u}_{k-1}}$ & measurement index of all sensors in a single slot (Section~\ref{subsec:sol_I1}) \\
        & $\iota = 1,\ldots,m$ & feature index \\
        \hline
        \multirow{3}{*}{Markov Chain} & $\mathrm{X}_k$ & MC state in slot $k$ \\
        & $e_i$ & a generic MC state (e.g., sit, walk, run, etc.) \\
        & $n$ & MC size \\
        \hline
        \multirow{3}{*}{Policy} & $N_{\rm tot}$ & maximum number of measurements in a single slot \\
        & $N_s^{\mathbf{u}_{k-1}}$ & number of measurements of sensor $s$ using the policy $\mathbf{u}_{k-1}$ \\
        & $\mathbf{u}_{k-1}$ & decision applied in slot $k$ \\
        \hline
        \multirow{3}{*}{Statistics} & $\mathrm{y}_{u,s,k},\mathrm{z}_{u,s,k}$ & $u$-th measurement of sensor $s$ in slot $k$ at the sensor and FC side, respectively\\
        & $\mathrm{h}_{u,s,k},\hat{\mathrm{h}}_{u,s,k}$ & real and estimate channel gain of the $u$-th tx measurement of sensor $s$ in slot $k$ \\
        & $c_{\iota,s,k}$ & $\iota$-th feature of sensor $s$ in slot $k$ \\
        \hline
        \multirow{2}{*}{Errors} & $\mathrm{w}_{\rm ch}$ & channel estimation error \\
        & $\mathrm{w}_{\rm noise}$ & channel AWGN noise \\
        \bottomrule
    \end{tabular}
\end{table*}

We study a system composed of $S$ sensors which track an unknown underlying system and transmit their data to a common fusion center. Time is slotted and the state of the system in slot $k$, $\mathrm{X}_k$, follows a Markov evolution according to a transition probability matrix $\mathbf{T}$ of size $n\times n$. The state $\mathrm{X}_k$ assumes values in the set $\{e_1,\ldots,e_n\}$ (e.g., $e_1 = $sit, $e_2 = $run, etc.). At every time step, sensor $s = 1,\ldots,S$ measures a feature related to the current state of the system. The measurement is noisy and follows a normal distribution $\mathcal{N}(m_{s,i},Q_{s,i})$, where $m_{s,i}$ and $Q_{s,i}$ are the mean and variance of the feature measured by sensor $s$ when the underlying state of the system is $e_i$~\cite{Thatte2011}. Since the features are state dependent, we exploit them to track the underlying system.

In a single time slot, sensor $s$ extracts $N_s^{\mathbf{u}_{k-1}}$ measurements (or samples) according to the centralized decision $\mathbf{u}_{k-1}$ made by the fusion center in the previous time slot. We denote by $\mathbf{u}_{k-1}$ the column vector with entries $\mathbf{u}_{k-1} = [N_1^{\mathbf{u}_{k-1}},\ldots,N_S^{\mathbf{u}_{k-1}}]^{\rm T}$.
The maximum number of samples extracted in a single time slot is $N_{\rm tot} \geq \sum_{s = 1}^S N_s^{\mathbf{u}_{k-1}}$.
We assume that the $N_s^{\mathbf{u}_{k-1}}$ measurements are statistically independent and identically distributed within the same slot, but the model may be extended as in~\cite{Thatte2011} to consider correlation between different samples.

The set $\mathbf{y}_k = \{\mathrm{y}_{u,s,k}, \forall u = 1,\ldots,N_s^{\mathbf{u}_{k-1}}, \forall s=1,\ldots,S\}$ represents all measurements collected by all sensors in time slot $k$ and $u$ is the index of a measurement in a single time slot. The set $\mathbf{y}_k$ can be seen as the realization of a random variable $\mathbf{Y} = [\mathrm{Y}_1,\ldots,\mathrm{Y}_S]$ (throughout, random variables are denoted with capital letters). 
The probability density function (pdf) of $\mathbf{Y}$ when the state of the system $e_i$ and the number of samples per sensor $\mathbf{u}_{k-1}$ are given, is denoted by (we adopt a notation similar to~\cite{Zois2013})~
\begin{align}\label{eq:f_Y}
    f_{\mathbf{Y}}(\mathbf{y}_k | e_i,\mathbf{u}_{k-1}) = \prod_{s = 1}^S \prod_{u = 1}^{N_s^{\mathbf{u}_{k-1}}} f_{\mathrm{Y}_{s}}(\mathrm{y}_{u,s,k} | e_i),
\end{align}

\noindent where we exploited the independence of the measurements among different sensors and over time, given the underlying state of the system.
In the following, we consider the notation indicated in Table~\ref{tab:parameters}.

\subsection{Channel} \label{subsec:channel}

Every measurement $\mathrm{y}_{u,s,k}$ is sent over a wireless communication link to the fusion center. The FC receives a channel distorted and noisy version of the measurement~
\begin{align}
    \mathrm{z}_{u,s,k} = \mathrm{h}_{u,s,k} \mathrm{y}_{u,s,k} + \mathrm{w}_{\rm noise},
\end{align}

\noindent where $\mathrm{w}_{\rm noise}$ is a realization of a random variable $\mathrm{W}_{\rm noise} \sim \mathcal{N}(0,\sigma_{\rm noise}^2)$, and $\mathrm{h}_{u,s,k}$ denotes the channel gain.

We assume that only a noisy version of $\mathrm{h}_{u,s,k}$ is available at the FC, namely $\hat{\mathrm{h}}_{u,s,k}$, defined as follows~\cite[Section II]{Ubaidulla2011}~
\begin{align}\label{eq:delta_h}
    \mathrm{h}_{u,s,k} = \hat{\mathrm{h}}_{u,s,k} + \mathrm{w}_{\rm ch},
\end{align}

\noindent where $\mathrm{W}_{\rm ch} \sim \mathcal{N}(0,\sigma_{\mathrm{h}}^2)$ summarizes the channel estimation errors. The received signal at the receiver side is~
\begin{align}
    &\mathrm{z}_{u,s,k} = \hat{\mathrm{h}}_{u,s,k} \mathrm{y}_{u,s,k} + \mathrm{y}_{u,s,k} \mathrm{w}_{\rm ch} + \mathrm{w}_{\rm noise}, \label{eq:z_usk}\\
    &\mathrm{Z}_s = \hat{\mathrm{h}}_{u,s,k} \mathrm{Y}_s + \mathrm{Y}_s \mathrm{W}_{\rm ch} + \mathrm{○W}_{\rm noise}, \label{eq:Z_s}
\end{align}

\noindent where Equation~\eqref{eq:Z_s} is expressed in the random variables domain.

For estimation purposes, since the FC only knows the term $\mathrm{z}_{u,s,k}$ (and \emph{not} $\mathrm{y}_{u,s,k}$), we need to compute its pdf. Two approaches can be considered.

%TODO CHECK REAL COMPLEX GAUSSIANS

\noindent \textbf{Non-robust Design.} The simplest choice is to neglect $\mathrm{w}_{\rm ch}$ in Equation~\eqref{eq:z_usk}. When the state of the system, $e_i$, and the estimate channel gain, $\hat{\mathrm{h}}_{u,s,k}$, are given, the resulting pdf is~
\begin{align}
    \mathrm{Z}_{s} | e_i, \hat{\mathrm{h}}_{u,s,k} \sim \mathcal{N}(\hat{\mathrm{h}}_{u,s,k} m_{s,i},\hat{\mathrm{h}}_{u,s,k}^2 Q_{s,i} + \sigma_{\rm noise}^2).
\end{align}

\noindent \textbf{Robust Design.} When $\mathrm{w}_{\rm ch}$ is explicitly taken into account, the pdf of $\mathrm{Z}_s$ given $e_i$ and $\hat{\mathrm{h}}_{u,s,k}$ is not Gaussian anymore. In this case, we have to consider the product of two Gaussian random variables, $\mathrm{Y}_{s}$ and $\mathrm{W}_{\rm ch}$. First, note that $\mathrm{Y}_s | e_i \sim \mathcal{N}(m_{s,i},Q_{s,i})$ can be decomposed as $m_{s,i} + \mathrm{Y}_s^0 | e_i$, with $\mathrm{Y}_s^0 \sim \mathcal{N}(0,Q_{s,i})$. Thus the product $\mathrm{Y}_s \mathrm{W}_{\rm ch}$ given $e_i$ can be written as~
\begin{align}\label{eq:Y_W_plus_Y0}
    \mathrm{Y}_s \mathrm{W}_{\rm ch} | e_i &= m_{s,i} \mathrm{W}_{\rm ch} + \mathrm{Y}_s^0 \mathrm{W}_{\rm ch} | e_i.
\end{align}

\noindent The pdf of the second term $\mathrm{Y}_s^0 \mathrm{W}_{\rm ch} | e_i$ is~
\begin{align}\label{eq:f_YW}
    f_{\mathrm{Y}_s\mathrm{W}_{\rm ch}}(x | e_i) = \frac{K_0\Big(\frac{|x|}{\sqrt{Q_{s,i}}\sigma_{\rm ch}}\Big)}{\pi \sqrt{Q_{s,i}} \sigma_{\rm ch}},
\end{align}

\noindent where $K_0(\cdot)$ is the modified Bessel function of the second kind. The pdf of $\mathrm{Z}_{s}$ can be written as the convolution of Equation~\eqref{eq:f_YW} and a Gaussian pdf with mean $\hat{\mathrm{h}}_{u,s,k}m_{s,i}$ and variance $\hat{\mathrm{h}}_{u,s,k}^2 Q_{s,i} + m_{s,i}^2 \sigma_{\rm ch}^2 + \sigma_{\rm noise}^2$, and, in general, is not Gaussian.
\\

While the non-robust approach is simpler because it only involves Gaussian random variables, it is less accurate than the robust one. An intermediate solution consists of using a modified robust approach obtained neglecting the term $\mathrm{Y}_s^0 \mathrm{W}_{\rm ch}$ in~\eqref{eq:Y_W_plus_Y0}, which can be shown to have a small impact on $\mathrm{Y}_s$. We adopt the intermediate approach in the remainder of the paper, so that using a Gaussian framework is still possible. The channel estimation error is taken into account via the term $m_{s,i} \mathrm{W}_{\rm ch}$:~
\begin{align}\label{eq:Z_s_distro}
    \mathrm{Z}_{s} | e_i, \hat{\mathrm{h}}_{u,s,k} \sim \mathcal{N}(\hat{\mathrm{h}}_{u,s,k} m_{s,i},\hat{\mathrm{h}}_{u,s,k}^2 Q_{s,i} + m_{s,i}^2\sigma_{\rm ch}^2 + \sigma_{\rm noise}^2).
\end{align}

\section{Tracking}\label{sec:tracking}

At the end of time slot $k$, the fusion center knows the sequence $\mathcal{F}_k$, defined as~
\begin{align}
	\mathcal{F}_k = \{\mathcal{Z}_k,\hat{\mathcal{H}}_k\},
\end{align}

\noindent where $\mathcal{Z}_k \triangleq \{\mathbf{z}_{0},\ldots,\mathbf{z}_{k}\}$ is the temporal sequence of the received measurements and $\hat{\mathcal{H}}_k \triangleq \{\hat{\mathbf{h}}_{0},\ldots,\hat{\mathbf{h}}_{k}\}$.
The goal of the system is to track the underlying hidden Markov process (i.e., estimate $\mathrm{X}_k$ in every time slot $k$). Towards this goal, we exploit the sequence $\mathcal{F}_k$ as follows.

At the end of every time slot, we update the \emph{belief} of the state of the system defined as~
\begin{align}\label{eq:p_k_k_x}
	\mathbf{p}_{k|k} \triangleq [\mathbb{P}(\mathrm{X}_k = e_1 | \mathcal{F}_k), \ldots, \mathbb{P}(\mathrm{X}_k = e_n | \mathcal{F}_k)],
\end{align}

\noindent which represents the estimated probability of observing $e_1,\ldots,e_n$ at the FC. To determine~\eqref{eq:p_k_k_x}, we exploit the sensor measurements as well as the channel observations. The belief $\mathbf{p}_{k|k}$ can be optimally evaluated as\footnote{For ease of notation, in the following we use $\mathbb{P}(\cdot)$ also to refer to probability density functions.}~
\begin{align}\label{eq:p_k_k_Bayes}
	\mathbb{P}(\mathrm{X}_k = e_i | \mathcal{F}_k) = \frac{\mathbb{P}(\mathrm{X}_k = e_i, \mathcal{F}_k | \mathcal{F}_{k-1})}{\mathbb{P}(\mathcal{F}_k | \mathcal{F}_{k-1})}.
\end{align}

\noindent The denominator can be computed with the following sum~
\begin{align}
	&\mathbb{P}(\mathcal{F}_k | \mathcal{F}_{k-1}) = \sum_{\ell = 1}^n \mathbb{P}(\mathrm{X}_k = e_\ell,\mathcal{F}_k|\mathcal{F}_{k-1}). \label{eq:denominator}
\end{align}

\noindent Therefore, since every term of the sum in~\eqref{eq:denominator} has the same form of the numerator in~\eqref{eq:p_k_k_Bayes}, we only focus on $\mathbb{P}(\mathrm{X}_k = e_i, \mathcal{F}_k|\mathcal{F}_{k-1})$. By definition, we have $\mathcal{F}_k = \{\mathcal{Z}_k,\hat{\mathcal{H}}_k\} = \mathcal{F}_{k-1} \cup \{\mathbf{z}_{k},\hat{\mathbf{h}}_{k}\}$, therefore~
\begin{subequations}
\label{eq:P_x_h_z_hat_h_u}
\begin{align}
    \mathbb{P}(\mathrm{X}_k = e_i, \mathcal{F}_k|\mathcal{F}_{k-1}) &= \mathbb{P}(\mathrm{X}_k = e_i,  \mathbf{Z} = \mathbf{z}_{k}, \hat{\mathbf{H}} = \hat{\mathbf{h}}_{k} | \mathcal{F}_{k-1}) \\
        & = \mathbb{P}(\mathbf{Z} = \mathbf{z}_{k} | \mathrm{X}_k = e_i, \hat{\mathbf{H}} = \hat{\mathbf{h}}_{k}, \mathcal{F}_{k-1}) \label{eq:P_x_h_z_hat_h_u1} \\
        & \ \times \mathbb{P}(\hat{\mathbf{H}} = \hat{\mathbf{h}}_{k} | \mathrm{X}_k = e_i, \mathcal{F}_{k-1}) \label{eq:P_x_h_z_hat_h_u2}\\
        & \ \times  \mathbb{P}(\mathrm{X}_k = e_i | \mathcal{F}_{k-1}). \label{eq:P_x_h_z_hat_h_u3}
\end{align}
\end{subequations}

\begin{figure}[!t]
  \centering
  \includegraphics[trim = 0mm 0mm 0mm 0mm, width=1\columnwidth]{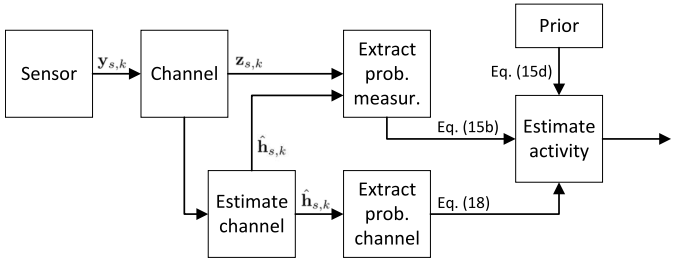}
  \caption{Exact estimation scheme (with $\mathbf{y}_{s,k} \triangleq [\mathrm{y}_{1,s,k},\ldots,\mathrm{y}_{N_s^{\mathbf{u}_{k-1}},s,k}]$ and similarly for $\mathbf{z}_{s,k}$ and $\hat{\mathbf{h}}_{s,k}$).}
  \label{fig:ori}
\end{figure}

\noindent We now analyze the three terms~\eqref{eq:P_x_h_z_hat_h_u1}-\eqref{eq:P_x_h_z_hat_h_u3} separately (see \figurename~\ref{fig:ori} for a graphical representation of the estimation scheme).

\textbf{First Term~\eqref{eq:P_x_h_z_hat_h_u1}.} First, we remark that the policy $\mathbf{u}_{k-1}$ is decided in slot $k-1$ and applied in slot $k$. Given the sequence $\mathcal{F}_{k-1}$, $\mathbf{u}_{k-1}$ is uniquely determined (i.e., $\mathbf{u}_{k-1}$ is a deterministic function of $\mathcal{F}_{k-1}$). The map between $\mathcal{F}_{k-1}$ and $\mathbf{u}_{k-1}$ is discussed in Subsection~\ref{subsec:MDP}.

Consider now~\eqref{eq:P_x_h_z_hat_h_u1}. Given the underlying state of the system, $\mathbf{Z}$ depends upon $\mathcal{F}_{k-1}$ only through $\mathbf{u}_{k-1}$. Similar to Equation~\eqref{eq:f_Y}, and using the results of the robust design in Section~\ref{subsec:channel}, we obtain~
\begin{align}\label{eq:f_Z_usk}
    \begin{split}
    &\mathbb{P}(\mathbf{Z} = \mathbf{z}_{k} | \mathrm{X}_k = e_i, \hat{\mathbf{H}} = \hat{\mathbf{h}}_{k}, \mathcal{F}_{k-1})= \prod_{s = 1}^S \prod_{u = 1}^{N_s^{\mathbf{u}_{k-1}}} \!\!\! f_{\mathrm{Z}_{s}}(\mathrm{z}_{u,s,k} | e_i, \hat{\mathrm{h}}_{u,s,k}),
    \end{split}
\end{align}

\noindent where $f_{\mathrm{Z}_{s}}(\mathrm{z}_{u,s,k} | e_i, \hat{\mathrm{h}}_{u,s,k})$ is a Gaussian random variable defined in~\eqref{eq:Z_s_distro}.

\textbf{Second Term~\eqref{eq:P_x_h_z_hat_h_u2}.} The channel itself can be used to track the underlying state of the system. We discuss how to approximate the term $\mathbb{P}(\hat{\mathbf{H}} = \hat{\mathbf{h}}_{k} | \mathrm{X}_k = e_i, \mathcal{F}_{k-1})$ in Subsection~\ref{subsec:channel_appr}.

\textbf{Third Term~\eqref{eq:P_x_h_z_hat_h_u3}.}
For the last factor in~\eqref{eq:P_x_h_z_hat_h_u}, we exploit the previous belief of the system and the total probability theorem:~
\begin{align}\label{eq:P_xk_GIV_Fkm1}
    &\mathbb{P}(\mathrm{X}_k = e_i | \mathcal{F}_{k-1}) = \sum_{j = 1}^n \mathbf{T}[j,i]  \mathbb{P}(\mathrm{X}_{k-1} = e_j| \mathcal{F}_{k-1}),
\end{align}

\noindent where the term $\mathbb{P}(\mathrm{X}_{k-1} = e_j| \mathcal{F}_{k-1})$ is a \emph{prior} and represents the belief in state $k-1$, whereas $\mathbf{T}[j,i]$ is the entry in position $(j,i)$ of the transition probability matrix $\mathbf{T}$ defined in Section~\ref{sec:system_model}.

Combining~\eqref{eq:p_k_k_Bayes}-\eqref{eq:P_xk_GIV_Fkm1}, $\mathbf{p}_{k|k}$ can be recursively computed starting from an initial belief of the system.

\subsection{Channel Approximation} \label{subsec:channel_appr}

Since the channel may exhibit temporal correlation within the same slot, modeling it as a whole random variable $\hat{\mathbf{H}}$ would be a hard task and would also incur in high numerical complexity.
Since our goal is to estimate the underlying state of the system, we are interested in deriving a simpler representation of the channel. %which is tractable and computationally feasible. 
In particular, we introduce an approximation which separates the realizations of the channel from its features, used to account for the temporal correlation. In general, given the sequence $\hat{h}_{1,s,k},\ldots,\hat{h}_{N_s^{\mathbf{u}_{k-1}},s,k}$, a lot of different features can be selected according to the particular scenario and depending upon the considered sensor (some of the most common ones are~\cite{Archasantisuk2015}: the Integrated Signal Level, the Mean Value, the Modified Mean Values, the Signal Square Integral, the Variance, the Root Mean Square, the Level Change, the Level Crossing, the Slope Change, the Willison Amplitude, the Histogram, the Range or the Slope of the Critical Point. We now discuss how to include the features in our model.

Define the exact probability of observing $\hat{\mathbf{H}}$ as~
\begin{align}
    \mathbb{P}_{\mathrm{h}} &\triangleq \mathbb{P}(\hat{\mathbf{H}} = \hat{\mathbf{h}}_{k} | \mathrm{X}_k = e_i, \mathcal{F}_{k-1}).
\end{align}

\noindent We approximate the probability $\mathbb{P}_{\mathrm{h}}$ as\footnote{In the remainder of the paper, we always use~\eqref{eq:P_h_approx} when we deal with the channel probabilities.}~
\begin{align}\label{eq:P_h_approx}
    \mathbb{P}_{\mathrm{h}} \approx \underbrace{\mathbb{P}(\hat{\mathbf{H}}^{\rm (i.i.d.)} = \hat{\mathbf{h}}_{k} | \mathrm{X}_k = e_i, \mathcal{F}_{k-1})}_{\rm \scriptstyle channel\ realizations} \underbrace{f_{\mathbf{C}}(\mathbf{c}_k | e_i,\mathbf{u}_{k-1})}_{\mathclap{\rm \scriptstyle temporal\ correlation}},
\end{align}

\noindent where the first term accounts for the channel realizations in a simple way, whereas the other term considers the channel temporal correlation (i.e., the features) and $\mathbf{u}_{k-1}$ is uniquely determined from $\mathcal{F}_{k-1}$. With~\eqref{eq:P_h_approx}, we can easily model the channel gain by studying two different quantities.

The random variable $\hat{\mathbf{H}}^{\rm (i.i.d.)}$ is the approximation of $\hat{\mathbf{H}}$ but, differently from the original version, it is assumed independent and identically distributed (i.i.d.) in every time slot and for every sensor. Therefore, its pdf can be simply expressed as (also in this case $\hat{\mathbf{H}}^{\rm (i.i.d.)}$ depends upon $\mathcal{F}_{k-1}$ only through $\mathbf{u}_{k-1}$)~
\begin{align}
    \mathbb{P}(\hat{\mathbf{H}}^{\rm (i.i.d.)} = \hat{\mathbf{h}}_{k} | \mathrm{X}_k = e_i, \mathcal{F}_{k-1}) = \prod_{s = 1}^S \prod_{u = 1}^{N_s^{\mathbf{u}_{k-1}}} f_{{\hat{\mathrm{H}}}_{s}^{\rm (i.i.d.)}}({\hat{\mathrm{h}}}_{u,s,k}|e_i).
\end{align}

\noindent Using Equation~\eqref{eq:delta_h}, the pdf of ${\hat{\mathrm{H}}}_{s}^{\rm (i.i.d.)}$ can be derived from $\mathrm{H}_{s}^{\rm (i.i.d.)}$ and $\mathrm{W}_{\rm ch}$. It has been shown that a Gamma or Weibull distribution is a good fit of $\mathrm{H}_{s}^{\rm (i.i.d.)}$ to experimental data collected for WBAN channels~\cite{Smith2011}.

\begin{figure}[!t]
  \centering
  \includegraphics[trim = 0mm 0mm 0mm 0mm, width=1\columnwidth]{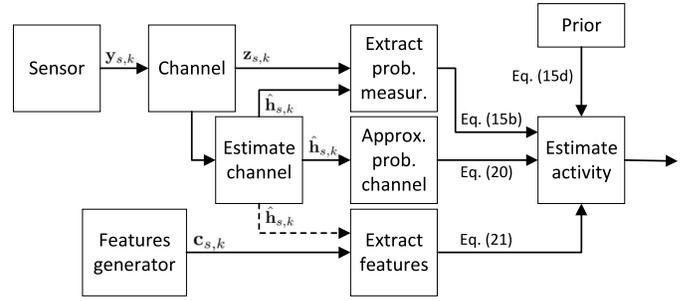}
  \caption{Approximate estimation scheme (with $\mathbf{y}_{s,k}$, $\mathbf{z}_{s,k}$ and $\hat{\mathbf{h}}_{s,k}$ as in \figurename~\ref{fig:ori} and $\mathbf{c}_{s,k} \triangleq [\mathrm{c}_{1,s,k},\ldots,\mathrm{c}_{m,s,k}]$). Formally, $\mathbf{c}_k$ should be derived from $\hat{\mathbf{h}}_k$ (dash-line). However, in our scheme we approximate it with a feature generation block.}
  \label{fig:mod}
\end{figure}

The second term of Equation~\eqref{eq:P_h_approx} represents the pdf of the features of the channel. The quantity $\mathbf{c}_k$ is the vector of features computed from $\hat{\mathbf{h}}_k$ in slot $k$. Since analyzing all the features is computational demanding, we only consider a subset of $m$ features. We identify the $\iota$-th feature of sensor $s$ in slot $k$ as $\mathrm{c}_{\iota,s,k}$, with $\iota = 1,\ldots,m$. The features are described by a joint random variable $\mathbf{C} = [C_{1,1},\ldots,C_{m,S}]$ ($C_{\iota,s}$ is the random variable associated with the $\iota$-th feature of sensor $s$), whose pdf $f_{\mathbf{C}}(\mathbf{c}_{k} | e_i,\mathbf{u}_{k-1})$, with $\mathbf{c}_{k} \triangleq \{\mathrm{c}_{\iota,s,k}, \forall \iota = 1,\ldots,m, \forall s = 1,\ldots,S\}$, can be evaluated empirically and is summarized by the ``features generator'' block in \figurename~\ref{fig:mod}. We consider independently distributed features, so that~
\begin{align}
    f_{\mathbf{C}}(\mathbf{c}_k | e_i,\mathbf{u}_{k-1}) = \prod_{s = 1}^S \prod_{\iota = 1}^{m} f_{\mathrm{C}_{\iota,s}}(\mathrm{c}_{\iota,s,k} | e_i, N_s^{\mathbf{u}_{k-1}}).
\end{align}

\noindent We expect that the higher the number of measurements per sensor $N_s^{\mathbf{u}_{k-1}}$, the smaller the variance of $\mathrm{C}_{\iota,s}$.\footnote{Iideally, $f_{\mathrm{C}_{\iota,s}}(\mathrm{c}_{\iota,s,k} | e_i, N_s^{\mathbf{u}_{k-1}})$ would become a single Dirac delta function when a lot of channel measurements were collected.}

%The pdfs $f_{\mathrm{C}_{\iota,s}}(\mathrm{c}_{\iota,s,k} | e_i, N_s^{\mathbf{u}_{k-1}})$ are known at the FC.

In summary, we rewrite~\eqref{eq:P_h_approx} as~
\begin{align}\label{eq:P_h_approx2}
    \begin{split}
        \mathbb{P}_{\mathrm{h}} = \prod_{s = 1}^S \bigg( \prod_{u = 1}^{N_s^{\mathbf{u}_{k-1}}} &f_{{\hat{\mathrm{H}}}_{s}^{\rm (i.i.d.)}}({\hat{\mathrm{h}}}_{u,s,k}|e_i) \\
        \prod_{\iota = 1}^m \hspace{2mm} &f_{\mathrm{C}_{\iota,s}}(\mathrm{c}_{\iota,s,k} | e_i, N_s^{\mathbf{u}_{k-1}}) \bigg).
    \end{split}
\end{align}

%TODO CHECK In the next we use $m = 1$ (i.e., only one feature) and omit the index $\iota$ for the sake of presentation simplicity, but the analysis can be easily extended to the generic case $m > 1$.

\section{Optimization} \label{sec:optimization}

The goal of the system is to simultaneously achieve high detection accuracy and low energy expenditure. These two conflicting objectives can be handled as a multi-objective weighted minimization problem. We define the instantaneous reward function as~
\begin{align}\label{eq:r_instant}
    r(\mathbf{p}_{k|k},\mathbf{u}_k) \triangleq (1-\lambda) \Delta(\mathbf{p}_{k|k}) + \lambda c(\mathbf{u}_k),
\end{align}

\noindent where $\Delta(\mathbf{p}_{k|k})$ represents the average estimation error, $c(\mathbf{u}_k)$ is an energy cost function increasing with $\mathbf{u}_k$ and $\lambda \in [0,1]$ is the weight.
We express $\Delta(\mathbf{p}_{k|k})$ as~
\begin{subequations}
\begin{align}
    \Delta(\mathbf{p}_{k|k}) &\triangleq \sum_{i = 1}^n \mathbb{E}\left[(x_{i,k} - \mathbb{P}(\mathrm{X}_k = e_i | \mathcal{F}_k))^2  | \mathcal{F}_k \right]\\
    & = 1 - \sum_{i = 1}^n \mathbb{P}(\mathrm{X}_k = e_i | \mathcal{F}_k)^2,
\end{align}
\end{subequations}

\noindent where the second equality can be derived after some algebraic manipulations.
While~\eqref{eq:r_instant} represents the instantaneous reward in a single slot, we are interested in the long-term optimization, thus, the long-run reward function becomes\footnote{$R_\mu$ can also be redefined using a discount factor instead of $\frac{1}{K}$ if the main focus is on the initial time slots. All our results can be straightforwardly extended to such a case.}~
\begin{align}
    R_\mu \triangleq \mathbb{E}\left[ \lim_{K \to \infty} \frac{1}{K}\sum_{k = 1}^K r(\mathbf{p}_{k|k},\mathbf{u}_k) \right].
\end{align}

\noindent The expectation is taken with respect to the measurements and to the channels. The policy $\mu = [\mathbf{u}_1,\mathbf{u}_2,\ldots]$ defines the number of samples gathered in every time slot for each sensor. The optimization problem is~
\begin{align} \label{eq:mu_star}
    \mu^\star = \argmin{\mu} \{ R_\mu \}.
\end{align}

\subsection{Markov Decision Process Formulation} \label{subsec:MDP}
The problem can be viewed as a Partially-Observable Markov Decision Process (POMDP)~\cite{Zois2013} and converted to an equivalent MDP~\cite{Puterman1995} for solution. The Markov Chain (MC) state of the converted problem is the belief $\mathbf{p}_{k|k}$ (it can be shown that this represents a sufficient statistic for control purposes~\cite{Kaelbling1998}) and a policy $\mu$ specifies the number of samples to gather and transmit for every possible combination of $\mathbf{p}_{k|k}$. Since we focus on long-term policies, we also drop all the dependencies from the slot index $k$ and enumerate the states of the Markov chain space with an index $\psi = 1,2,\ldots$ (i.e., the belief space is now composed of $\mathbf{p}_1,\mathbf{p}_2,\ldots$). Also, it can be shown that $\mu^\star$ is a deterministic policy (see~\cite[Theorems 6.2.9 and 6.2.10]{Puterman1995} for the discounted horizon case), thus we restrict our study to the class of deterministic strategies.

Common algorithms to solve average long-term MDPs are the Value Iteration Algorithm (VIA) or the Policy Iteration Algorithm (PIA)~\cite[Vol.~II, Sec.~4]{Bertsekas2005}. The basic step of both these approaches is the \emph{policy improvement step}, in which the following cost-to-go function is updated~
\begin{align}
    &J(\mathbf{p}_\psi) \leftarrow \min_{\mathbf{u}} \{ K(\mathbf{p}_\psi,\mathbf{u}) \}, \label{eq:J_k}\\
    &K(\mathbf{p}_\psi,\mathbf{u}) \triangleq \mathbb{E}_{\mathbf{Z},\hat{\mathbf{H}}}[ \underbrace{r(\mathbf{p}_\psi,\mathbf{u}) + J(\mathbf{p}_{\psi'})}_{\textstyle \triangleq (\bullet)} | \mathbf{p}_\psi,\mathbf{u}],\label{eq:K_k}
\end{align}

\noindent where $r(\mathbf{p}_\psi,\mathbf{u})$ represents the instantaneous reward defined in~\eqref{eq:r_instant}, whereas $J(\mathbf{p}_{\psi'})$ accounts for the future rewards. The index of the new state, given $\mathbf{Z}$, $\hat{\mathbf{H}}$, $\mathbf{p}_\psi$ and $\mathbf{u}$, is $\psi'$. The corresponding belief $\mathbf{p}_{\psi'}$ is derived as~
\begin{subequations}
\label{eq:p_psi_prime}
\begin{align}
    &\mathbf{p}_{\psi'}(i) = \frac{\mathbb{P}(\mathbf{Z} = \mathbf{z}, \hat{\mathbf{H}} = \hat{\mathbf{h}} | \mathrm{X} = e_i, \mathbf{u}) \mathbf{p}_\psi(i)}{\mathbb{P}(\mathbf{Z} = \mathbf{z}, \hat{\mathbf{H}} = \hat{\mathbf{h}} | \mathbf{p}_\psi, \mathbf{u})},\\
    &\mathbb{P}(\mathbf{Z} = \mathbf{z}, \hat{\mathbf{H}} = \hat{\mathbf{h}} | \mathbf{p}_\psi, \mathbf{u}) = \sum_{j = 1}^n \mathbb{P}(\mathbf{Z} = \mathbf{z}, \hat{\mathbf{H}} = \hat{\mathbf{h}} | \mathrm{X} = e_j, \mathbf{u}) \mathbf{p}_\psi(j),
\end{align}
\end{subequations}

\noindent where $\mathbf{p}_\psi(i)$ is the $i$-th entry of $\mathbf{p}_\psi$ and $\mathrm{X}$ is the state of the underlying system without time index. Note that $\sum_{i = 1}^n \mathbf{p}_{\psi'}(i) = 1$. Equation~\eqref{eq:p_psi_prime} is the equivalent of \eqref{eq:p_k_k_Bayes} in the long term and can be evaluated as described in Section~\ref{sec:tracking}.

The expectation in Equation~\eqref{eq:K_k} is taken with respect to the measurements $\mathbf{Z}$ and the channel conditions $\hat{\mathbf{H}}$ and can be rewritten by definition as~
\begin{align} \label{eq:expectation}
    &\mathbb{E}_{\mathbf{Z},\hat{\mathbf{H}}}[(\bullet) | \mathbf{p}_\psi,\mathbf{u}] = \int \int (\bullet) \mathbb{P}(\mathbf{Z} = \mathbf{z}, \hat{\mathbf{H}} = \hat{\mathbf{h}} | \mathbf{p}_\psi, \mathbf{u}) \ \mbox{d}\mathbf{z} \ \mbox{d}\hat{\mathbf{h}}.
\end{align}

Note that with our formulation, the instantaneous reward $(1-\lambda) \Delta(\mathbf{p}_\psi) + \lambda c(\mathbf{u})$ in Equation~\eqref{eq:K_k} does not depend upon $\mathbf{Z}$ or $\hat{\mathbf{H}}$, and thus can be moved outside the expectation term:~
\begin{align}
    \begin{split} \label{eq:K_k_revised}
        &K(\mathbf{p}_\psi,\! \mathbf{u}) =  (1-\lambda) \Delta(\mathbf{p}_\psi) \! +\! \lambda c(\mathbf{u})\! +\! \mathbb{E}[J(\mathbf{p}_{\psi'}) | \mathcal{F}_k].
    \end{split}
\end{align}

In summary, from~\eqref{eq:J_k}-\eqref{eq:K_k_revised}, we can rewrite the policy improvement step as~
\begin{subequations}\label{eq:problem}
\begin{align}
    \min_{\mathbf{u}} \{\lambda c(\mathbf{u})\! +\! \mathbb{E}[J(\mathbf{p}_{\psi'}) | \mathcal{F}_k]\}
\end{align}
\vspace{-\belowdisplayskip}
\vspace{-\abovedisplayskip}
\begin{alignat}{2}
\shortintertext{s.t.:}
    & \eqref{eq:p_psi_prime}, \qquad \eqref{eq:expectation} , \qquad \mathbf{u} = [N_1^{\mathbf{u}},\ldots, N_S^{\mathbf{u}}]^{\rm T}, \\
    & N^{\mathbf{u}} \triangleq \sum_{s = 1}^S N_s^{\mathbf{u}} \leq N_{\rm tot}, \qquad N_s^{\mathbf{u}} \geq 0, \qquad  \forall s = 1,\ldots,S.
\end{alignat}
\end{subequations}

The constraint $N^{\mathbf{u}} \leq N_{\rm tot}$ was introduced in Section~\ref{sec:system_model}. The constraint $N_s^{\mathbf{u}} \geq 0$ ensures that we have a non-negative number of measurements.
Note that, because of the term $c(\mathbf{u})$ (which is a non-decreasing function of $N^{\mathbf{u}}$) in the objective function, the constraint $N^{\mathbf{u}} \leq N_{\rm tot}$ is not satisfied with equality, in general (a particular case in which the equality holds is for $\lambda = 0$).

\section{Analysis} \label{sec:analysis}

Determining the \emph{optimal} solution described in the previous section requires a challenging numerical evaluation.
In particular, there are two main issues involved:
\begin{itemize}
    \item I1) \eqref{eq:problem} must be solved for every belief~\cite{Zhou2001};
    \item I2) minimizing Problem~\eqref{eq:problem} is computationally demanding. In particular, the optimization involves a combinatorial problem with multi-dimensional integrals.
\end{itemize}

In this section, we propose solutions to both of these problems. In particular, I1) is handled in Subsection~\ref{subsec:sol_I1}, in which we introduce bounds which can be efficiently computed on a small subset of the belief space, whereas we deal with I2) in Subsection~\ref{subsec:sol_I2} by reducing the complexity multi-dimensional integrals and introducing a sub-optimal scheme.

\subsection{Concavity Properties}

We first introduce some preliminary results on the concavity properties of the reward function.
\begin{thm}\label{thm:K_conc}
    Consider a set of $n$ beliefs $\mathbf{b} = [\mathbf{b}^{(1)},\ldots,\mathbf{b}^{(n)}]$ such that a generic belief $\mathbf{p}_\psi$ can be written as~
    \begin{align}\label{eq:p_k_k_v_i_b_k_k}
        &\mathbf{p}_\psi = \sum_{i = 1}^n \alpha_i \mathbf{b}^{(i)},\\
        &\sum_{i = 1}^n \alpha_i = 1, \quad \alpha_i \geq 0,\quad \forall i = 1,\ldots,n
    \end{align}
    
    \noindent where $\alpha_i$ is a constant. Then, function $K(\cdot)$ is lower bounded by~
    \begin{align} \label{eq:K_concave}
        K(\mathbf{p}_\psi,\mathbf{u}) \geq r(\mathbf{p}_\psi,\! \mathbf{u}) \! +\! \sum_{i = 1}^n \alpha_i (K(\mathbf{b}^{(i)},\mathbf{u}) \! -\! r(\mathbf{b}^{(i)},\! \mathbf{u})).
    \end{align}
    
    \begin{proof}
        See \appendixname~\ref{proof:lower_bound}.
    \end{proof}
\end{thm}

Several different techniques for defining the vector $\mathbf{b}$ can be found in the literature~\cite{Shani2013} (e.g., in our numerical evaluation we will apply Levejoy's grid method~\cite{Lovejoy1991}). A consequence of Theorem~\ref{thm:K_conc} is derived in the following corollary.
\begin{corol}\label{corol:J_concave}
    The function~
    \begin{align}
        J(\mathbf{p}_\psi) - (1-\lambda)\Delta(\mathbf{p}_\psi)
    \end{align}
    
    \noindent is concave in every entry of~$\mathbf{p}_\psi$.
    \begin{proof}
        See \appendixname~\ref{proof:J_concave}.
    \end{proof}
\end{corol}

Note that the cost term $c(\mathbf{u})$ is not included in the previous corollary since it does not depend upon $\mathbf{p}_\psi$.

\subsection{Issue I1)} \label{subsec:sol_I1}

Using the results of Corollary~\ref{corol:J_concave}, we introduce a lower and an upper bound to the reward $R_{\mu^\star}$. In Section~\ref{sec:numerical_results} we will show that these bounds are tight and can be used as approximation to $R_{\mu^\star}$.

\subsubsection{Lower Bound} \label{subsubsec:lower_bound}

From Corollary~\ref{corol:J_concave}, we have ($\alpha_i$ and $\mathbf{b}^{(i)}$ are defined in Theorem~\ref{thm:K_conc}):~
\begin{align} \label{eq:J_concave}
    J(\mathbf{p}_\psi) \geq (1-\lambda)\Delta(\mathbf{p}_\psi) \!+\! \sum_{i = 1}^n \alpha_i (J(\mathbf{b}^{(i)}) \!-\! (1 \!-\! \lambda)\Delta(\mathbf{b}^{(i)})).
\end{align}

In general, two different beliefs $\mathbf{p}_{\psi^\prime}$ and $\mathbf{p}_{\psi^{\prime\prime}}$ can be written as a combination of different vectors $\mathbf{b}^\prime$ and $\mathbf{b}^{\prime\prime}$, respectively. We denote by $\mathcal{B}$ the set which contains all the vectors $\mathbf{b}^\prime$, $\mathbf{b}^{\prime\prime}$, etc., so that \emph{every} belief $\mathbf{p}_\psi$ can be written as a linear combination of the elements of $\mathcal{B}$.\footnote{We remark that, even if the minimum size of $\mathcal{B}$ is $n$, it may be more convenient to use $|\mathcal{B}| > n$.}
A lower bound to the optimal performance can be obtained as follows. The optimal $J(\mathbf{p}_\psi)$ is only computed in all the elements of $\mathcal{B}$, whereas, in all other states, $J(\mathbf{p}_\psi)$ is approximated with the right-hand side of~\eqref{eq:J_concave}. Using this approximation at every step of the value iteration algorithm~\cite{Bertsekas2005}, we obtain a lower bound to $R_{\mu^\star}$, denoted by $\tilde{R}_{\mu^\star}$. Note that, while the original $J(\mathbf{p}_\psi)$ should be computed in an infinite state space, $\mathcal{B}$ is a finite set and its size can be defined according to the desired precision.

\begin{comment}

\subsubsection{Loose Lower Bound} \label{subsubsec:loose_lower_bound}

An approach similar to the one proposed in the previous subsection can be found in the literature~\cite{TODO}. However, in general the approximation is performed by exploiting the concavity of $J(\mathbf{p}_\psi)$ is also concave in $\mathbf{p}_\psi$. Using this information instead of Corollary~\ref{corol:J_concave}, similar results can be found and the ``loose lower bound'' of \figurename~\ref{fig:up_low} can be derived.

\end{comment}

\subsubsection{Upper Bound} \label{subsubsec:upper_bound}

\begin{figure}[t]
  \centering
  \includegraphics[trim = 0mm 0mm 0mm 0mm,  clip, width=.9\columnwidth]{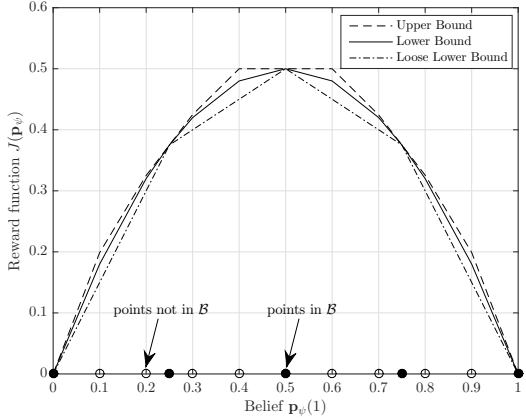}
  \caption{Lower and upper bounds to the cost-to-go function when $n = 2$ (number of MC states) for a single iteration of the value iteration algorithm. The loose lower bound is found as $J(\mathbf{p}_\psi) \geq \sum_{i = 1}^n \alpha_i J(\mathbf{b}^{(i)})$.}
  \label{fig:up_low}
\end{figure}

Using~\eqref{corol:J_concave}, it is also possible to derive an upper bound to $R_{\mu^\star}$. In particular, since $J(\mathbf{p}_\psi) - (1-\lambda)\Delta(\mathbf{p}_\psi)$ is concave, it is upper bounded by a piece-wise linear function composed of its tangent curves (see \figurename~\ref{fig:up_low} for a graphic interpretation)~\cite{Krishnamurthy2002}. While, in the general case, the tangents can be computed in arbitrary points of the belief space, it becomes numerically easier to derive them for the values in $\mathcal{B}$, so that lower and upper bounds can be simultaneously evaluated.

\subsection{Bounds-Based Policy} \label{subsec:BBP}

So far, we have described only techniques for computing bounds to the long-term reward $R_{\mu^\star}$ and in particular we introduced $\tilde{R}_{\mu^\star}$. However, while computing $R_{\mu^\star}$ requires us to evaluate the optimal policy $\mu^\star$ in all the belief space, when we compute $\tilde{R}_{\mu^\star}$, $\mu^\star$ is specified only in the subset of beliefs $\mathcal{B}$. We formally denote this difference by writing $R_{\mu^\star}$ and $\tilde{R}_{\mu_{\mathcal{B}}^\star}$. In general, given $\mu_{\mathcal{B}}^\star$, we do not know the optimal policy in all the remaining states. Therefore, since we need to explicitly define a policy for every belief, we extend $\mu_{\mathcal{B}}^\star$ with a probabilistic approach and derive the Bounds-Based Policy (BBP) as follows (we denote $\mathbf{u}_\psi^\star$ as the solution of~\eqref{eq:problem} when the belief is $\mathbf{p}_\psi$):~
\begin{align}\label{eq:prob_policy}
    \mathbb{P}(\mathbf{U} = \mathbf{u} | \mathbf{p}_\psi) = \begin{cases}
        \delta_{\mathbf{u},\mathbf{u}_\psi^\star}, \quad &\mbox{if } \mathbf{p}_\psi \in \mathcal{B}, \\
        \sum_{i = 1}^n \alpha_i\delta_{\mathbf{u},\mathbf{u}_\psi^\star}, \quad &\mbox{otherwise},
    \end{cases} \quad \mbox{(BBP)}
\end{align}

\noindent where $\alpha_i$ and $\mathbf{b}^{(i)}$ are defined as in Theorem~\ref{thm:K_conc}, $\mathbf{U}$ is the policy random variable and $\delta_{\cdot,\cdot}$ is the Kronecker delta function. In the first case, if $\mathbf{p}_\psi \in \mathcal{B}$, with probability equal to $1$, $\mathbf{U}$ assumes the value $\mathbf{u}_\psi^\star$, which is the solution of~\eqref{eq:problem} and is enclosed in $\mu_{\mathcal{B}}^\star$. Otherwise, the second case can be obtained as a combination of the policies computed in $\mathcal{B}$ and it is always less than~$1$. We will further discuss BBP in the numerical evaluation. Note that multiple choices of $\alpha_i$ may exist. In our case, we define them using Lovejoy's strategy~\cite{Lovejoy1991}.

\subsection{Issue I2)} \label{subsec:sol_I2}

Issue I2) is composed of two steps which we analyze sequentially in Subsections~\ref{subsubsec:multi_dim_integrals} and~\ref{subsubsec:iterative_opt}.

\subsubsection{Multi-Dimensional Integrals} \label{subsubsec:multi_dim_integrals}

In order to simplify I2), we need to separate the multi-dimensional integrals. This can be done optimally as follows. First, focus on the measurements in a single slot and enumerate every measurement from $1$ to $N^{\mathbf{u}} = \sum_{s = 1}^S N_s^{\mathbf{u}}$ with an index $\nu$ (i.e., $\mathrm{z}^{(1)},\ldots,\mathrm{z}^{(N^{\mathbf{u}})}$) such that there is a one-to-one mapping with each pair $(u,s)$. The term $\mathrm{z}^{(\nu)}$ coincides with $\mathrm{z}_{u,s}$ (with the Markov formulation we neglect the temporal index $k$). The pdf of $\mathrm{z}^{(\nu)}$ given the underlying state of the system $e_i$ and the underlying estimate channel $\hat{h}^{(\nu)}$ is $f_{\mathrm{Z}_s}(\mathrm{z}^{(\nu)} | e_i, \hat{h}^{(\nu)})$ (defined according to Section~\ref{subsec:channel}).

\begin{thm} \label{thm:belief_simplified}
    Given $\mathbf{p}_\psi$, the new belief $\mathbf{p}_{\psi'}$ can be equivalently computed as~
    \begin{align}\label{eq:p_k_k_recur}
        &\mathbf{p}_{\psi'}(i) = \frac{\mathbb{P}(\hat{\mathbf{H}} = \hat{\mathbf{h}} | \mathrm{X} = e_i, \mathbf{u}) \mathbf{p}_{\psi'}(i|N^{\mathbf{u}})}{\sum_{j = 1}^n \mathbb{P}(\hat{\mathbf{H}} = \hat{\mathbf{h}} | \mathrm{X} = e_j, \mathbf{u}) \mathbf{p}_{\psi'}(j|N^{\mathbf{u}})},
    \end{align}
    
    \noindent where $\mathbf{p}_{\psi'}(i | \nu)$ is recursively defined as\footnote{Note that $\mathbf{p}_{\psi'}(i | \nu)$ implicitly depends upon $\hat{h}^{(1)},\ldots,\hat{h}^{(\nu)}$.}~
    \begin{align}\label{eq:x_j_F_kv}
        \begin{split}
            &\mathbf{p}_{\psi'}(i | \nu) \triangleq  \begin{cases}
                \!\!\frac{\textstyle f_{\mathrm{Z}_s}(\mathrm{z}^{(\nu)} | e_i, \hat{h}^{(\nu)}) \mathbf{p}_{\psi'}(i|\nu-1)}{\textstyle\sum_{j = 1}^n f_{\mathrm{Z}_s}(\mathrm{z}^{(\nu)} | e_j, \hat{h}^{(\nu)}) \mathbf{p}_{\psi'}(j|\nu-1)},  \!\!\!\! & \mbox{if } \nu \geq 1, \\
                \mathbf{p}_{\psi}(i), \!\!\!\! & \mbox{if } \nu = 0.
            \end{cases}
        \end{split}
    \end{align}

    \begin{proof}
        See \appendixname~\ref{proof:belief_simplified}.
    \end{proof}
\end{thm}

With the previous theorem, instead of considering all the measurements together, we iteratively compute a partial belief for every new measurement $\nu$ exploiting the old partial belief at stage $\nu-1$ as in Equation~\eqref{eq:x_j_F_kv}. This allows us to decompose the $N^{\mathbf{u}}$-dimensional integral of the measurements in Equation~\eqref{eq:expectation} in $N^{\mathbf{u}}$ separate uni-dimensional integrals \textit{without} performance loss.

So far, Theorem~\ref{thm:belief_simplified} has been presented as a method to simplify the integral of the measurements. However, when we approximate $\mathbb{P}_{\mathrm{h}}$ as in Equation~\eqref{eq:P_h_approx2}, then the same technique can be extended to the channel to simplify the corresponding integral. This is possible because the temporal correlation of the channel has been entirely contained in a separate random variable $\mathbf{C}$. We give the extension of Theorem~\ref{thm:belief_simplified} in the next corollary.

\begin{corol}\label{corol:belief_simplified}
    Given $\mathbf{p}_\psi$, the new belief $\mathbf{p}_{\psi'}$ can be equivalently computed as~
    \begin{align}
        &\mathbf{p}_{\psi'}(i) = \frac{\mathbf{a}_{\psi'}(i|S)}{\sum_{j = 1}^n \mathbf{a}_{\psi'}(j|S)},
    \end{align}
        
    \noindent $\mathbf{a}_{\psi'}(i | s)$ is recursively defined as~
    \begin{align}
        \begin{split}
            &\mathbf{a}_{\psi'}(i | s) \triangleq  \begin{cases}
                \!\!\frac{\textstyle \prod_{\iota = 1}^{m} f_{\mathrm{C}_{\iota,s}}(\mathrm{c}_{\iota,s} | e_i, N_s^{\mathbf{u}}) \mathbf{a}_{\psi'}(i|s-1)}{\textstyle\sum_{j = 1}^n \prod_{\iota = 1}^{m} f_{\mathrm{C}_{\iota,s}}(\mathrm{c}_{\iota,s} | e_j, N_s^{\mathbf{u}}) \mathbf{a}_{\psi'}(i | s-1)},  \!\!\!\! & \mbox{if } s \geq 1, \\
                \mathbf{p}_{\psi'}(i | N^{\mathbf{u}}), \!\!\!\! & \mbox{if } s = 0
            \end{cases}
        \end{split}
    \end{align}
    
    \noindent and $\mathbf{p}_{\psi'}(i | \nu)$ is given in~\eqref{eq:x_j_F_kv} where we replace the terms $f_{\mathrm{Z}_s}(\mathrm{z}^{(\nu)} | e_i, \hat{h}^{(\nu)})$ with~
    \begin{align}
        f_{\mathrm{Z}_s,\mathrm{H}_s}(\mathrm{z}^{(\nu)},\hat{h}^{(\nu)} | e_i) = f_{\hat{\mathrm{H}}_s}(\hat{\mathrm{h}}^{(\nu)} | e_i) f_{\mathrm{Z}_s}(\mathrm{z}^{(\nu)} | e_i, \hat{h}^{(\nu)}).
    \end{align}
\end{corol}

With Corollary~\ref{corol:belief_simplified}, all the integrals in~\eqref{eq:expectation} can be reduced to simpler uni-dimensional integrals, so that I2) is partially solved.

\subsubsection{Iterative Optimization} \label{subsubsec:iterative_opt}

%TODO SEE COMMENT Mitra

% to find the combinations, write \sum_{s1 = 0}^N \sum_{s2 = 0}^{N-s1} \sum_{s3 = 0}^{N-s1-s2} 1

Another issue of I2) is related to the combinatorial nature of the problem. In particular, in~\eqref{eq:problem}, $\frac{\prod_{s=1}^{S-1} (N^{\mathbf{u}}+s)}{(S-1)!}$ possible combinations are available for a fixed $N^{\mathbf{u}}$. In order to reduce this number, we define a sub-optimal greedy approach which leads to local minima. The idea consists in dividing the $N^{\mathbf{u}}$ samples in smaller subsets and optimally allocate these subsets. In particular, define an ordered vector of $L$ integers $[N^{(1)},\ldots,N^{(L)}]$ such that $\sum_{\ell = 1}^L N^{(\ell)} = N^{\mathbf{u}}$. Then, starting with $N^{(1)}$, compute the optimal policy $\mathbf{u}^1$ (e.g., $\mathbf{u}^1 = [1, 0, 0]^{\rm T}$ with $S = 3$ and $N^{(1)} = 1$) and store it. Then, consider $N^{(1)}+N^{(2)}$ and find $\mathbf{u}^{(2)}$ by solving the problem using the partial policy $\mathbf{u}^{(1)}$ and optimizing only the choice of the remaining $N^{(2)}$ sensors to use (e.g., considering the previous example, the possible choices are $\mathbf{u}^{(2)} = [2, 0, 0]^{\rm T}, [1, 1, 0]^{\rm T}, [1, 0, 1]^{\rm T}$ with $N^{(2)} = 1$). Repeat the procedure for every $\ell = 1,\ldots,L$. At the last step, the policy is fully specified by $\mathbf{u} = \mathbf{u}^{(L)}$.

\section{Numerical Results} \label{sec:numerical_results}

\begin{figure}[t]
  \centering
  \includegraphics[trim = 0mm 0mm 0mm 0mm,  clip, width=.9\columnwidth]{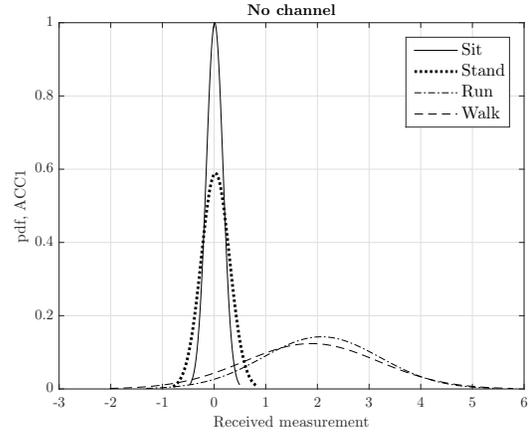}
  \caption{Normalized measurement pdfs $f_{\mathrm{Z}_{\rm ACC1}}(\mathrm{z} | e_i, \hat{\mathrm{h}}) = f_{\mathrm{Y}_{\rm ACC1}}(\mathrm{y} | e_i)$ of the first sensor ACC1 for different states $e_i \in \{\mbox{sit,stand,run,walk}\}$. For ACC1, no channel is considered.}
  \label{fig:z_distro_ACC1}
\end{figure}
\begin{figure*}[t]
  \centering
  \includegraphics[trim = 15mm 0mm 30mm 0mm,  clip, width=.9\textwidth]{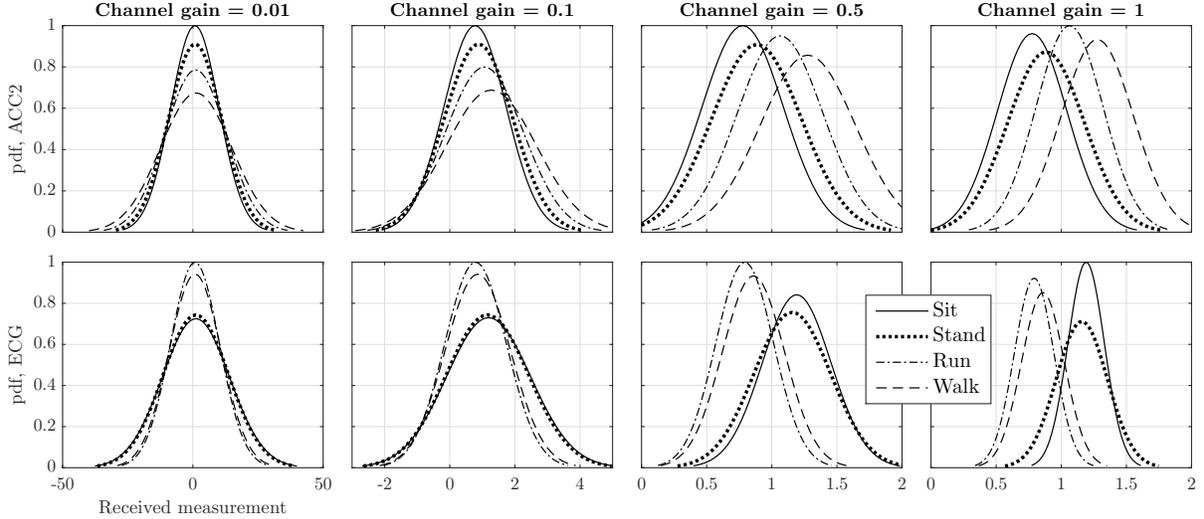}
  \caption{Normalized measurement pdfs $f_{\mathrm{Z}_{\rm ACC2}}(\mathrm{z} | e_i,\hat{\mathrm{h}})$ and  $f_{\mathrm{Z}_{\rm ECG}}(\mathrm{z} | e_i, \hat{\mathrm{h}})$ of the second and third sensors (ACC2 and ECG), respectively, for different states $e_i \in \{\mbox{sit,stand,run,walk}\}$ and values of the estimate channel $\hat{\mathrm{h}} \in \{0.01,0.1,0.5,1\}$. When the channel gain is low, it becomes more difficult to distinguish between the different states because the received signal is more degraded.}
  \label{fig:z_distro_ACC2_ECG}
\end{figure*}

\begin{figure}[t]
  \centering
  \includegraphics[trim = 0mm 0mm 0mm 0mm,  clip, width=.9\columnwidth]{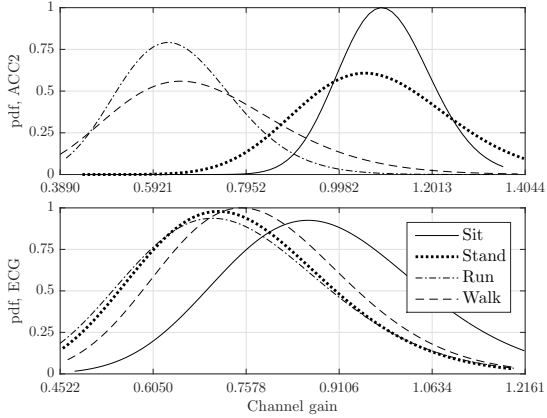}
  \caption{Normalized real channel pdfs $f_{\mathrm{H}_{{\rm ACC2}}^{\rm (i.i.d.)}}(\mathrm{h}| e_i)$ and  $f_{\mathrm{H}_{{\rm ECG}}^{\rm (i.i.d.)}}(\mathrm{h}| e_i)$ of the second and third sensors (ACC2 and ECG), respectively, for different states $e_i \in \{\mbox{sit,stand,run,walk}\}$.}
  \label{fig:h_distro}
\end{figure}
\begin{figure*}[t]
  \centering
  \includegraphics[trim = 15mm 0mm 30mm 0mm,  clip, width=.9\textwidth]{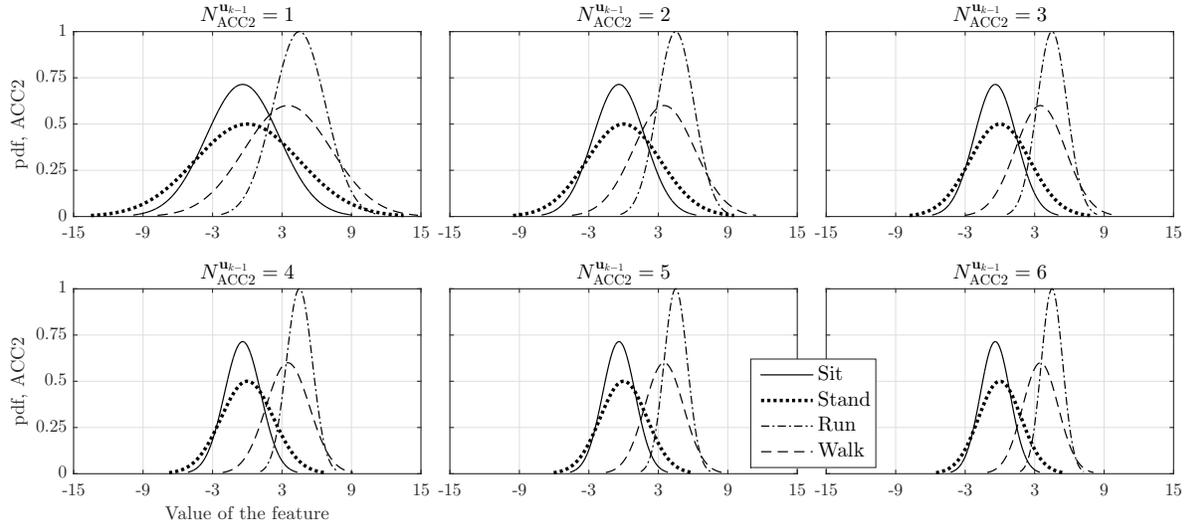}
  \caption{Normalized feature pdfs $f_{\mathrm{C}_{1,{\rm ACC2}}}(\mathrm{c} | e_i, N_{\rm ACC2}^{\mathbf{u}_{k-1}})$ of the second sensor ACC2 for different states $e_i \in \{\mbox{sit,stand,run,walk}\}$ and number of samples extracted $N_{\rm ACC2}^{\mathbf{u}_{k-1}} \in 1,\ldots,6$. The higher $N_{\rm ACC2}^{\mathbf{u}_{k-1}}$, the easier to distinguish among different states.}
  \label{fig:Pk_distro}
\end{figure*}

We consider a WBAN based on the KNOW-ME system described in~\cite{Thatte2011} composed of two accelerometers (ACC1 and ACC2) and an electrocardiography sensor (ECG), which track the current activity of a subject (sitting, standing, running, walking). Different costs are associated with the data reception by different sensors: ACC1 is located inside the fusion center (which, e.g., can be a mobile phone in a pocket), thus the reception of a sample from ACC1 is energy efficient, whereas ACC2 and ECG are on-body sensors and the data reception from these requires more energy. ACC2 provides good quality measurements, but also experiences bad channel conditions (e.g., it is a sensor on the wrist, subject to movements and thus its channel changes significantly over time), thus it is likely that its measurements are highly corrupted by noise. In our example, the feature extracted from ACC2 can be interpreted as a measure of the temporal periodicity of the movement of the arm where the sensor is placed on. The electrocardiography can be placed in the chest, thus it experiences less movements than ACC2 and is characterized by a better channel. We assume that no features are extracted from ECG.
Finally, since ACC1 is located inside the fusion center, no communication channel is considered for this sensor.

\textbf{Parameters.} If not otherwise stated, we use the following parameters~\cite{Thatte2011}: up to $N_{\rm tot} = 6$ measurements per slot, $n = 4$ states of the system (sitting, standing, running, walking), $S = 3$ sensors (ACC1, ACC2, ECG), a transition probability matrix~
\begin{align}
    \mathbf{T} = 
    \kbordermatrix{ & {\rm sit} & {\rm stand} & {\rm run} & {\rm walk} \\
        & 0.6 & 0.1 & 0   & 0.3 \\
        & 0.2 & 0.4 & 0.1 & 0.3 \\
        & 0   & 0.1 & 0.3 & 0.6 \\
        & 0.4 & 0   & 0.3 & 0.3 }
\end{align}

\noindent and a cost function $c(\mathbf{u}_k) = \sum_{s \in \{{\rm ACC1}, {\rm ACC2}, {\rm ECG}\}} \delta_s N_s^{\mathbf{u}_k}$, with $\boldsymbol{\delta} = [0.58, 0.776, 1]$ and $|\mathcal{B}| = 35$. The channel is characterized by $\sigma_{\rm ch} = 0.05$, $\sigma_{\rm noise} = 0.05$ and the effects of different path losses are already enclosed in the pdf $f_{\mathrm{H}_{s}^{\rm (i.i.d.)}}(\mathrm{h}| e_i)$. A single feature is considered (i.e., $m = 1$) and is extracted from ACC2. The densities of the measurements, channels and features are represented in \figurename s~\ref{fig:z_distro_ACC1}-\ref{fig:Pk_distro} (as in~\cite{Thatte2011,Smith2011} we used Gaussian pdfs for the measurements and Gamma pdfs for the channels). \figurename~\ref{fig:z_distro_ACC1} is referred to ACC1 and thus is not influenced by any channel value (the generated measurements are always perfectly ``received'' by the FC). Note that all the pdfs are Gaussian-distributed with different means and variances. In this case, it is almost impossible to distinguish between sitting/standing or running/walking. Instead, for the sensors ACC2 and ECG, as can be seen in~\figurename~\ref{fig:z_distro_ACC2_ECG}, the worse the channel, the more degraded the received signal and thus the more difficult to distinguish the four densities.\footnote{When the channel gain is $1$ and $\sigma_{\rm ch} = 0.05$, then there is an uncertainty of $5\%$ in the channel estimation. As the channel gain decreases, the relative uncertainty increases.} Note that all the densities are Gaussian distributed even when the channel gets worse. \figurename~\ref{fig:h_distro} shows the channel pdfs for the different states of the system. For ACC2, in the ``run'' and ``walk'' states, the average channel gains are lower than for the others, so it is more likely to receive degraded measurements. Finally, the distributions of the feature of ACC2 are represented in \figurename~\ref{fig:Pk_distro}. When more samples are extracted from the same sensor (i.e., higher $N_{{\rm ACC2}}^{\mathbf{u}_{k-1}}$), it becomes easier to distinguish the underlying state of the system. From the previous discussion, we can conclude that a lot of different trade-offs exist, thus determining, a priori, which is the best sensor to use is not an easy task.

\textbf{Bounds.} First, we discuss the bounds we derived in Section~\ref{subsec:sol_I1} for a simple case. For numerical tractability, we set $n = 2$ (``sit'' and ``stand''), $\mathbf{T}[{\rm sit},{\rm stand}] = 6/9$, $\mathbf{T}[{\rm stand},{\rm sit}] = 1/2$, $\sigma_{\rm ch} = 0.001$, $N_{\rm max} = 3$ and all the other parameters as defined before. Since, theoretically, the belief space is continuous, it is not possible to find numerically the \emph{true} optimal reward $R_{\mu^\star}$, and a discrete approximate approach with a sufficiently large number of quantization levels is required. Therefore, for computing $R_{\mu^\star}$, we used Levejoy's grid method~\cite{Lovejoy1991} with $200$ levels (i.e., $201$ beliefs), whereas for the upper and lower bounds we reduced the levels to $5$ (i.e., $|\mathcal{B}| = 6$ in Section~\ref{subsubsec:lower_bound}). Thus, all the previous $201$ beliefs can now be written as a linear combination of $6$ beliefs as defined in Theorem~\ref{thm:K_conc}.

\begin{figure}[!t]
  \centering
  \includegraphics[trim = 0mm 0mm 0mm 6mm,  clip, width=.9\columnwidth]{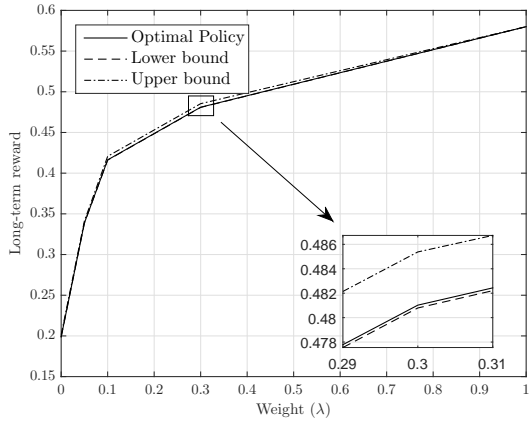}
  \caption{Long-term reward as a function of the weight $\lambda$ when $n = 2$, $\sigma_{\rm ch} = 0.001$ and $N_{\rm max} = 3$.}
  \label{fig:upper_lower}
\end{figure}

We generated our results for different values of the weighting $\lambda \in [0,1]$ and computed the corresponding long-term reward (see \figurename~\ref{fig:upper_lower}). It can be noted that the bounds are extremely tight in all the region and that the lower bound approximates the optimal strategy very well. Because of this, in the following we will always approximate the optimal reward $R_{\mu^\star}$, with its lower bound computed using Levejoy's approximation with a lower amount of beliefs, denoted by $\tilde{R}_{\mu^\star}$.\footnote{According to Subsection~\ref{subsec:BBP}, we remark that we should formally write $R_{\mu^\star}$ and $\tilde{R}_{\mu_{\mathcal{B}}^\star}$ (i.e., when computing the lower bound the policy is computed only in a subset of the belief space). In the following, we omit the subscript for notation clarity.}

\begin{figure}[!t]
  \centering
  \includegraphics[trim = 0mm 0mm 0mm 6mm,  clip, width=.9\columnwidth]{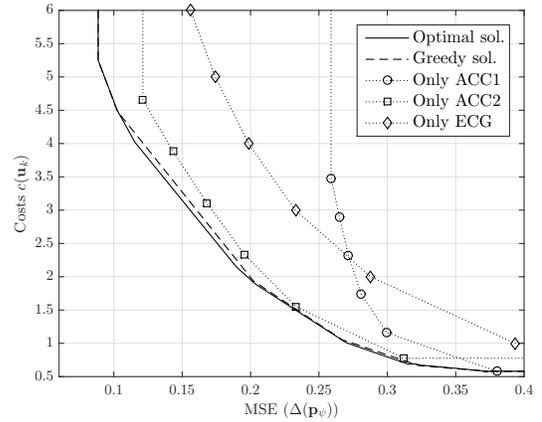}
  \caption{Lower bounds of the cost function as a function of the MSE evaluated optimally (continuous curve) or with a greedy approach (dash-curve). The marked-curves represent the performance of the policies which always use the same sensor.}
  \label{fig:sub_opt}
\end{figure}

\textbf{Sub-optimal strategies.} We now compare $\tilde{R}_{\mu^\star}$ with other strategies. \figurename~\ref{fig:sub_opt} shows the boundaries of the Pareto regions for the different techniques (we artificially lengthen the beginning of each curve for graphical comparison). The axes are the energy cost $c(\mathbf{u}_k)$ (defined in~\eqref{eq:r_instant}) and the probability of incorrect state prediction. The points with lower (higher) MSE (which is strictly related to the prediction error probability) are obtained for lower (higher) values of the weight $\lambda$. We consider the lower bound to the optimal policy $\tilde{R}_{\mu^\star}$, its greedy approximation, and the reward of three policies obtained by choosing the same sensor all the time. The curve referred to $\mu^\star$ always dominates the others, since the $\min$ operation in~\eqref{eq:problem} is solved optimally. We also plot $\tilde{R}_{\rm greedy}$, in which we used the approximation described in Section~\ref{subsubsec:iterative_opt} with $N^{(1)} = \ldots = N^{(L)} = 1$ for solving the $\min$ in~\eqref{eq:problem}. It is interesting to note that, even if the greedy policy uses a simpler approach to solve~\eqref{eq:problem}, it achieves almost optimal performance. Moreover, it is extremely easy to compute, thus it may be considered as a good alternative to $\mu^\star$. We also remark that the performance of the greedy approach can be further improved by choosing larger $N^{(m)}$ (see Section~\ref{subsubsec:iterative_opt}). Finally, we derived the three policies which use the same sensor all the time. In general, these are strongly sub-optimal, especially if ACC1 or ECG are considered. Since it is also possible to extract a feature from ACC2, choosing it provides better performance, even if it does not achieve $\tilde{R}_{\mu^\star}$.

\begin{figure}[!t]
  \centering
  \includegraphics[trim = 0mm 0mm 0mm 6mm,  clip, width=.9\columnwidth]{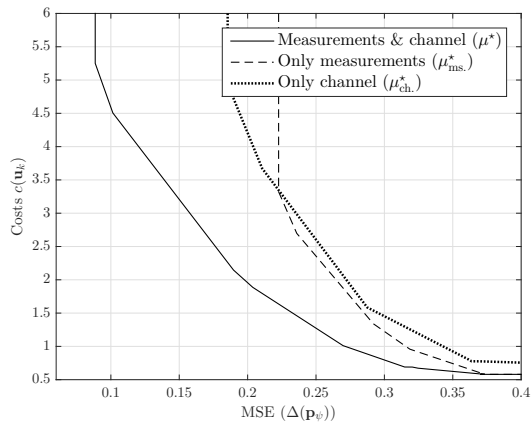}
  \caption{Lower bounds of the cost function as a function of the MSE for three different strategies $\mu^\star$, $\mu^\star_{\rm ms}$, $\mu^\star_{\rm ch}$.}
  \label{fig:strategy}
\end{figure}

\begin{figure}[!t]
  \centering
  \includegraphics[trim = 20mm 0mm 20mm 3mm, width=1\columnwidth]{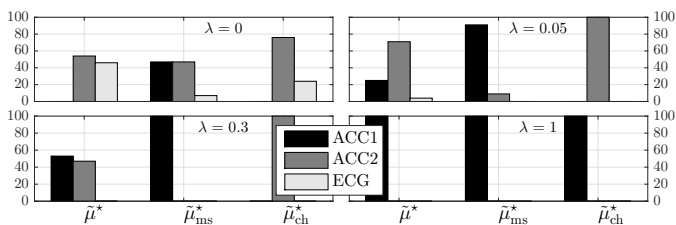}
  \caption{Percentages of time dedicated to every sensor for increasing values of the weight $\lambda$ and three different strategies $\mu^\star$, $\mu^\star_{\rm ms}$, $\mu^\star_{\rm ch}$.}
  \label{fig:bars}
\end{figure}

\textbf{Measurements and channel.} In this paragraph, we want to describe the importance of using measurements and the channel state jointly to obtain high performance (see \figurename s~\ref{fig:strategy} and~\ref{fig:bars}). In addition to $\mu^\star$, we introduce the strategies $\mu_{\rm ms}^\star$ and $\mu_{\rm ch}^\star$, which indicate the optimal policies obtained by using the measurements only and the channels only, respectively. Thus, for example, when $\mu_{\rm ms}^\star$ is used, even if the channels are available, they are not used for improving the tracking performance. \figurename~\ref{fig:strategy} shows the boundaries of the Pareto regions for the three cases. The main difference among these can be noted when the prediction error probability is the dominating term (upper left-side of the figure). In this region, $\mu_{\rm ms}^\star$ and $\mu_{\rm ch}^\star$ cannot reach MSEs as low as $\mu^\star$, which encloses both the advantages of the other two schemes. Instead, if the energy cost is the dominating factor, the three approaches are similar since using the most energy efficient sensors is sufficient to achieve energy efficiency.

%TODO As an example, by looking at \figurename s~\ref{fig:h_distro} and~\ref{fig:Pk_distro}, it can be seen that, using the channel only, it is not possible distinguish all the four states of the system, thus it shows how the channel cannot be used alone to determine the state of the system.

An additional comparison of the three schemes is shown in \figurename~\ref{fig:bars}, which represents the average usage time of every sensor for different values of $\lambda$. When $\lambda = 1$, the system chooses $N_{\rm ACC1}^{\mathbf{u}} = 1$, $N_{\rm ACC2}^{\mathbf{u}} = 0$ and $N_{\rm ECG}^{\mathbf{u}} = 0$ all the time so that to consume only $\delta_{\rm ACC1}$ in every slot (we remind that $\delta_{\rm ACC1}$ is the lowest among the three costs). Also, since the only concern is the energy cost, all policies coincide in this case. However, for $\lambda = 0$, the three schemes behave significantly differently. With $\mu^\star$, the first sensor ACC1 is never used since it does not have any communication channel to exploit as additional source of information and the quality of its measurements is comparable with the others. Similarly, $\mu_{\rm ch}^\star$ does not use ACC1 because there is no channel and its tracking system is based on the channel only. Finally, $\mu_{\rm ms}^\star$ uses all the sensors with different percentages in order to exploit and combine the measurements of every sensor. For the other values of $\lambda$, we obtain intermediate situations between the two. In particular, as $\lambda$ increases, the role of ACC1 becomes dominant until the situation in which ACC1 is used all the time.


\begin{figure}[!t]
  \centering
  \includegraphics[trim = 0mm 0mm 0mm 6mm,  clip, width=.9\columnwidth]{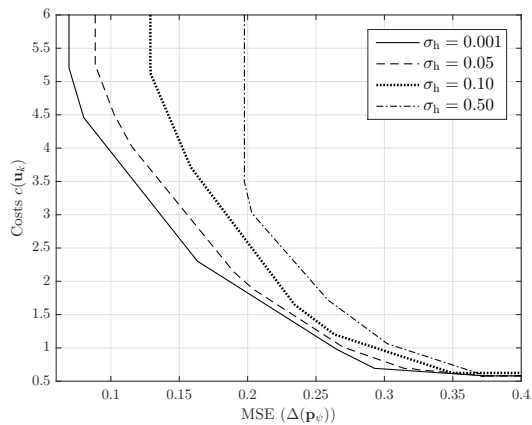}
  \caption{Lower bounds of the cost function as a function of the MSE for different values of the channel estimation error standard deviation $\sigma_{\rm ch}$.}
  \label{fig:sigma_h}
\end{figure}

\textbf{Channel errors.} Changing the estimation error $\sigma_{\rm ch}$ significantly affects the tracking performance. When $\sigma_{\rm ch}$ increases, in addition to obtain a worse estimation of the channel, the received measurements are also degraded, and thus they become less informative. This can be clearly seen in~\figurename~\ref{fig:z_distro_ACC2_ECG}, as previously explained. Note that, when $\sigma_{\rm ch}$ increases significantly, using ACC1 becomes the only choice to infer information about the underlying system. \figurename~\ref{fig:sigma_h} shows the boundary of the Pareto regions for different $\sigma_{\rm ch}$. When $\sigma_{\rm ch} = 0.001$, the estimation errors are almost negligible, thus this is a lower bound to the performance of the system. For higher $\sigma_{\rm ch}$, the MSE quickly increases, making the system unusable.

\begin{figure}[!t]
  \centering
  \includegraphics[trim = 0mm 0mm 0mm 6mm,  clip, width=.9\columnwidth]{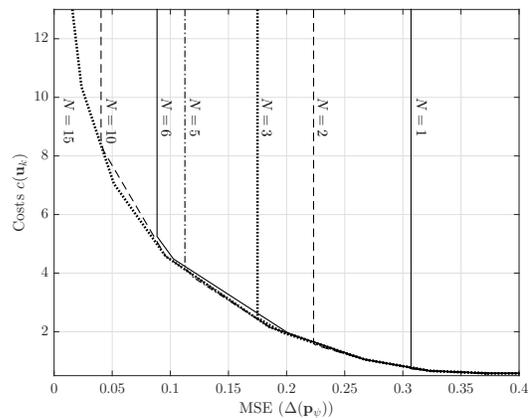}
  \caption{Lower bounds of the cost function as a function of the MSE for different values of the maximum number of samples per slot $N_{\rm max}$.}
  \label{fig:change_N}
\end{figure}

\textbf{Number of samples.} When $N_{\rm max}$ is small, the system performance is limited and reaching very low prediction error probabilities is not possible. In \figurename~\ref{fig:change_N}, we represent the boundaries of the Pareto regions for different values of $N_{\rm max}$ (as before, we lengthen the beginning of each curve for graphical purposes). Every case can be approximately seen as a sub-case of $N_{\rm max} = 15$ obtained for different values of $\lambda$. Indeed, the higher the value of $\lambda$, the lower the average number of used sensors per slot (we remark that even if $N_{\rm max}$ is fixed, a policy may use $N^{\mathbf{u}} \leq N_{\rm max}$ samples per slot, if necessary). It is interesting to note that the improvement obtained from $N_{\rm max}$ to $N_{\rm max}+1$ decreases with $N_{\rm max}$ (e.g., going from $N_{\rm max} = 1$ to $N_{\rm max} = 2$ provides a much larger improvement than going from $N_{\rm max} = 5$ to $N_{\rm max} = 6$). However, the energy costs increase linearly with the number of measurements, therefore a very high energy cost must be accrued to achieve low MSEs.

\begin{figure}[!t]
  \centering
  \includegraphics[trim = 0mm 0mm 0mm 6mm,  clip, width=.9\columnwidth]{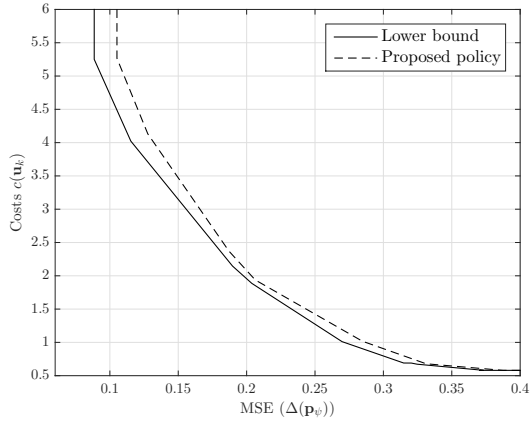}
  \caption{Lowe bound of the cost function as a function of the MSE and performance of BBP. Note that the continuous curve is a lower bound, whereas the other is achievable by BBP.}
  \label{fig:simul}
\end{figure}

\textbf{Probabilistic policy.} Finally, we discuss the bounds-based policy we introduced in Equation~\eqref{eq:prob_policy}. For clarity, we remark that $\tilde{R}_{\mu_{\mathcal{B}}^\star} = \tilde{R}_{\mu^\star} \leq R_{\mu^\star} \leq R_{\rm BBP}$. The first equality is by definition, the first inequality comes from Subsection~\ref{subsubsec:lower_bound} and the last inequality is by definition of the optimal policy. Nevertheless, BBP is computed using $\mu_{\mathcal{B}}^\star$, which is the only policy (partially) available when we compute $\tilde{R}_{\mu_{\mathcal{B}^\star}}$. However, differently from the lower and upper bounds, BBP is a policy which can implemented in a real system. \figurename~\ref{fig:simul} shows the boundaries of the Pareto regions of the lower bound and of BBP. It can seen that the two are quite close in a large region. However, if very strict constraints are imposed on the accuracy (abscissa values), then it may be necessary to compute the optimal scheme, which in turn may require a much higher computational cost.

%Therefore, also BBP can be seen as an upper bound to the optimal reward,

\section{Conclusions} \label{sec:conclusions}
We set up an active sensing problem in which sensor measurements and communication channel characteristics are used jointly to improve the system performance. Energy costs at the fusion center and estimation quality are handled together so as to characterize the Pareto regions of the system. We set up a POMDP framework and converted it to an equivalent MDP for solving. Exploiting the structural properties of the model we reduced the problem complexity and bounded the system performance. In particular, we decomposed the tracking update formula in a subset of simpler tasks which can be easily handled and, moreover, we used  the concavity properties of the cost-to-go function to introduce bounds to the long-term reward function and define a probabilistic sub-optimal policy. We built our numerical evaluation according to the specifications found in the literature and we proved the importance of considering the channel as an additional source of information in a WBAN scenario.

\section{Acknowledgments}
This research has been funded in part by the following grants and organizations: 
AFOSR FA9550-12-1-0215,
NSF CNS-1213128,
NSF CCF-1410009,
NSF CPS-1446901, 
ONR N00014-09-1-0700,
ONR N00014-15-1-2550, and Fondazione Ing. Aldo Gini.

\appendices

\section{Proof of Theorem~\ref{thm:K_conc}}\label{proof:lower_bound}

We first introduce the following proposition.

\begin{propos} 
For every step $I = 1,2\ldots$ of the value iteration algorithm~\cite[Vol.~II, Sec.~4.3.1]{Bertsekas2005},~
\begin{align} \label{eq:proof_K_Im1}
    K^{(I)}\left( \mathbf{p}_\psi \circ \frac{[a_1,\ldots,a_n]}{\sum_{j = 1}^n \mathbf{p}_\psi(j)a_j}, \mathbf{u} \right) \sum_{j = 1}^n \mathbf{p}_\psi(j)a_j
\end{align}

\noindent is concave in $\mathbf{p}_\psi$, where $a_j$ is a non-negative constant and $\circ$ is the Hadamard product.

\begin{proof}
The proof is by induction over the steps of the value iteration algorithm.
At step $I = 1$, we have~
\begin{align}
    K^{(1)}(\mathbf{p}_\psi,\mathbf{u}) = r(\mathbf{p}_\psi,\mathbf{u}) = (1-\lambda)\Delta(\mathbf{p}_\psi) + \lambda c(\mathbf{u}).
\end{align}

\noindent Since, on the right-hand side, only $\Delta(\cdot)$ depends upon $\mathbf{p}_\psi$, to prove the concavity we focus on the term $\Delta(\cdot)$:~
\begin{subequations}
\label{eq:proof_Delta_1}
\begin{align}
    &\Delta\left( \mathbf{p}_\psi \circ \frac{[a_1,\ldots,a_n]}{\sum_{j = 1}^n \mathbf{p}_\psi(j)a_j} \right)\sum_{j = 1}^n \mathbf{p}_\psi(j)a_j \\
    &= \Bigg(1-\sum_{i = 1}^n\left(\frac{\mathbf{p}_\psi(i)a_i}{\sum_{j = 1}^n \mathbf{p}_\psi(j)a_j}\right)^2\Bigg)\sum_{j = 1}^n \mathbf{p}_\psi(j)a_j \\
    &= \sum_{j = 1}^n \mathbf{p}_\psi(j)a_j - \frac{\sum_{j = 1}^n (\mathbf{p}_\psi(j)a_j)^2}{\sum_{j = 1}^n \mathbf{p}_\psi(j)a_j}.\label{eq:proof_Delta_1_3}
\end{align}
\end{subequations}

To prove that the previous term is concave, we compute its second order derivative with respect to $\mathbf{p}_\psi(i)$:~
\begin{align}
    \frac{\partial^2 \eqref{eq:proof_Delta_1_3}}{\partial \mathbf{p}_\psi(i)^2} = -2 a_i \frac{ \left(\sum_{j = 1 \atop j \neq i}^n a_j \mathbf{p}_\psi(j)\right)^2 + \sum_{j = 1 \atop j \neq i}^n \left(a_j \mathbf{p}_\psi(j)\right)^2}{\left(\sum_{j = 1}^n a_j \mathbf{p}_\psi(j)\right)^3},
\end{align}

\noindent which is always smaller than or equal to zero, thus~\eqref{eq:proof_K_Im1} holds for $I = 1$. Now, assume that~\eqref{eq:proof_K_Im1} holds for a generic $I-1$. At step $I$ we have~
\begin{align}
    K^{(I)}(\mathbf{p}_\psi,\mathbf{u}) = r(\mathbf{p}_\psi,\mathbf{u}) + \mathbb{E}_{\mathbf{Z},\hat{\mathbf{H}}}[J^{(I-1)}(\mathbf{p}_{\psi'}) | \mathbf{p}_\psi, \mathbf{u}].
\end{align}

\noindent We consider the two terms separately. The first term $r(\mathbf{p}_\Psi,\mathbf{u})$ coincides with $K^{(1)}(\cdot)$ and thus is concave when evaluated at $\frac{\mathbf{p}_\psi(i)a_i}{\sum_{j = 1}^n \mathbf{p}_\psi(j)a_j}$ and multiplied by $\sum_{j = 1}^n \mathbf{p}_\psi(j)a_j$. The second term can be expressed as in Equation~\eqref{eq:expectation}~
\begin{align} \label{eq:proof_E_J_Im1}
    \begin{split}
        &\mathbb{E}[J^{(I-1)}(\mathbf{p}_{\psi'}) | \mathbf{p}_\psi, \mathbf{u}] = \mathbb{E}[\min_{\mathbf{u}'} \{ K^{(I-1)}(\mathbf{p}_{\psi'}, \mathbf{u}') \} | \mathbf{p}_\psi, \mathbf{u}] \\
        & = \int \min_{\mathbf{u}'} \{ K^{(I-1)}(\mathbf{p}_{\psi'}, \mathbf{u}') \} \mathbb{P}(\mathbf{Z} = \mathbf{z}, \hat{\mathbf{H}} = \hat{\mathbf{h}} | \mathbf{p}_\psi, \mathbf{u}) \ \mbox{d}\mathbf{z} \ \mbox{d}\hat{\mathbf{h}},
    \end{split}
\end{align}

\noindent where $\mathbf{p}_{\psi'}$ is derived in Equation~\eqref{eq:p_psi_prime}. Note that the term $\mathbb{P}(\mathbf{Z} = \mathbf{z}, \hat{\mathbf{H}} = \hat{\mathbf{h}} | \mathbf{p}_\psi, \mathbf{u})$ can be moved inside the $\min$-operator. Using the inductive hypothesis and defining $a_i \triangleq \mathbb{P}(\mathbf{Z} = \mathbf{z}, \hat{\mathbf{H}} = \hat{\mathbf{h}} | \mathrm{X} = e_i, \mathbf{u})$, we have that every argument of the $\min$-operation is concave, thus~\eqref{eq:proof_E_J_Im1} is concave and the thesis is proved.
\end{proof}
\end{propos}

With the previous proposition, it is straightforward to also show that $K^{(I)}\left( \mathbf{p}_\psi, \mathbf{u} \right) - r(\mathbf{p}_\psi,\mathbf{u})$ is concave for every $I = 1,2,\ldots$, which is equivalent to~\eqref{eq:K_concave}.

\section{Proof of Corollary~\ref{corol:J_concave}}\label{proof:J_concave}

We want to prove that~
\begin{align}
    J(\mathbf{p}_\psi) \geq (1-\lambda)\Delta(\mathbf{p}_\psi) \!+\! \sum_{i = 1}^n \alpha_i (J(\mathbf{b}^{(i)}) \!-\! (1 \!-\! \lambda)\Delta(\mathbf{b}^{(i)})),
\end{align}

\noindent where $\mathbf{b}^{(i)}$ and $\alpha_i$ are defined as in Theorem~\ref{thm:K_conc}.
By definition,~
\begin{align}
    J(\mathbf{p}_\psi) \!-\! (1\!-\!\lambda)\Delta(\mathbf{p}_\psi) = \min_{\mathbf{u}} \{K(\mathbf{p}_\psi,\mathbf{u}) \!-\! (1\!-\!\lambda)\Delta(\mathbf{p}_\psi) \}
\end{align}

\noindent and using Theorem~\ref{thm:K_conc}, the right-hand side can be lower bounded by~
\begin{align}
    \begin{split}
    &\min_{\mathbf{u}} \{K(\mathbf{p}_\psi,\mathbf{u}) \!-\! (1\!-\!\lambda)\Delta(\mathbf{p}_\psi) \} \geq \min_{\mathbf{u}} \{r(\mathbf{p}_\psi,\mathbf{u}) \\
    &+\! \sum_{i = 1}^n\! \alpha_i (K(\mathbf{b}^{(i)},\!\mathbf{u}) \! -\! r(\mathbf{b}^{(i)},\! \mathbf{u})) \!-\! (1\!-\!\lambda)\Delta(\mathbf{p}_\psi) \}. 
    \end{split}\label{eq:proof_corol_2}
\end{align}

The terms $r(\mathbf{p}_\psi,\mathbf{u}) - (1\!-\!\lambda)\Delta(\mathbf{p}_\psi)$ can be reduced to $\lambda c(\mathbf{u})$ and the term $-\sum_{i = 1}^n \alpha_i r(\mathbf{b}^{(i)},\! \mathbf{u})$ can be simplified as~
\begin{align}
    -\sum_{i = 1}^n \alpha_i r(\mathbf{b}^{(i)},\! \mathbf{u}) = -\sum_{i = 1}^n \alpha_i (1-\lambda)\Delta(\mathbf{b}^{(i)}) - \lambda c(\mathbf{u})
\end{align}

\noindent because $\sum_{i = 1}^n \alpha_i = 1$ in Theorem~\ref{thm:K_conc}. Combining the previous expression and~\eqref{eq:proof_corol_2}, we obtain~
\begin{align}
    \begin{split}
        &J(\mathbf{p}_\psi) \!-\! (1\!-\!\lambda)\Delta(\mathbf{p}_\psi) \geq \min_{\mathbf{u}} \{\lambda c(\mathbf{u}) \\
        &+ \sum_{i = 1}^n \alpha_i (K(\mathbf{b}^{(i)},\mathbf{u}) \! -\! (1-\lambda)\Delta(\mathbf{b}^{(i)})) - \lambda c(\mathbf{u}) \},
    \end{split}
\end{align}

\noindent which coincides with~\eqref{eq:J_concave} and concludes the proof.

\section{Proof of Theorem~\ref{thm:belief_simplified}}\label{proof:belief_simplified}

First, we show by induction over $\nu$ that $\mathbf{p}_{\psi'}(i | \nu)$ defined as in~\eqref{eq:x_j_F_kv} is equivalent to (for notation simplicity we neglect the dependencies on the channel in the pdfs $f_{Z_s}(\cdot)$)~
\begin{align} \label{eq:proof_x_k_F_kv}
    &\bar{\mathbf{p}}_{\psi'}(i | \nu) = \frac{\textstyle \prod_{w=1}^{\nu} f_{\mathrm{Z}_s}(\mathrm{z}^{(w)} | e_i) \mathbf{p}_{\psi}(i)}{\textstyle\sum_{j = 1}^n \prod_{w=1}^{\nu} f_{\mathrm{Z}_s}(\mathrm{z}^{(w)} | e_j) \mathbf{p}_{\psi}(j)}.
\end{align}
        
For $\nu = 1$, \eqref{eq:x_j_F_kv} and~\eqref{eq:proof_x_k_F_kv} coincide. Assume that they coincide for a generic index $\nu > 1$. Then, for $\nu+1$  substitute~\eqref{eq:proof_x_k_F_kv} in~\eqref{eq:x_j_F_kv} and obtain~
\begin{subequations}
\begin{align}
    & \mathbf{p}_{\psi'}(i | \nu+1) = \frac{\textstyle f_{\mathrm{Z}_s}(\mathrm{z}^{(\nu+1)} | e_i) \mathbf{p}_{\psi'}(i|\nu)}{\textstyle\sum_{j = 1}^n f_{\mathrm{Z}_s}(\mathrm{z}^{(\nu+1)} | e_j) \mathbf{p}_{\psi'}(j|\nu)} \\
    &  = \frac{\textstyle f_{\mathrm{Z}_s}(\mathrm{z}^{(\nu+1)} | e_i) \bar{\mathbf{p}}_{\psi'}(i|\nu)}{\textstyle\sum_{j = 1}^n f_{\mathrm{Z}_s}(\mathrm{z}^{(\nu+1)} | e_j) \bar{\mathbf{p}}_{\psi'}(j|\nu)} \\
    & = \frac{\textstyle f_{\mathrm{Z}_s}(\mathrm{z}^{(\nu+1)} | e_i) \frac{\textstyle \prod_{w=1}^{\nu} f_{\mathrm{Z}_s}(\mathrm{z}^{(w)} | e_i) \mathbf{p}_{\psi}(i)}{\textstyle\sum_{\ell = 1}^n \prod_{w=1}^{\nu} f_{\mathrm{Z}_s}(\mathrm{z}^{(w)} | e_\ell) \mathbf{p}_{\psi}(\ell)}}{\textstyle\sum_{j = 1}^n f_{\mathrm{Z}_s}(\mathrm{z}^{(\nu+1)} | e_j) \frac{\textstyle \prod_{w=1}^{\nu} f_{\mathrm{Z}_s}(\mathrm{z}^{(w)} | e_j) \mathbf{p}_{\psi}(j)}{\textstyle\sum_{\ell = 1}^n \prod_{w=1}^{\nu} f_{\mathrm{Z}_s}(\mathrm{z}^{(w)} | e_\ell) \mathbf{p}_{\psi}(\ell)}} \nonumber\\
    & = \frac{\textstyle \prod_{w=1}^{\nu+1} f_{\mathrm{Z}_s}(\mathrm{z}^{(w)} | e_i) \mathbf{p}_{\psi}(i)}{\textstyle\sum_{j = 1}^n \prod_{w=1}^{\nu+1} f_{\mathrm{Z}_s}(\mathrm{z}^{(w)} | e_j) \mathbf{p}_{\psi}(j)} = \bar{\mathbf{p}}_{\psi'}(i | \nu+1).
\end{align}
\end{subequations}

\noindent Thus, we proved that \eqref{eq:x_j_F_kv}$\equiv$\eqref{eq:proof_x_k_F_kv}. 

Then, substitute~\eqref{eq:proof_x_k_F_kv} for $v = N^{\mathbf{u}}$ in~\eqref{eq:p_k_k_recur} to obtain~
\begin{align}\label{eq:proof_p_psi_prime}
    \mathbf{p}_{\psi'}(i) = \frac{\mathbb{P}(\hat{\mathbf{H}} = \hat{\mathbf{h}} | \mathrm{X} = e_i, \mathbf{u}) \prod_{v=1}^{N^{\mathbf{u}}} f_{\mathrm{Z}_s}(\mathrm{z}^{(v)} | e_i) \mathbf{p}_{\psi}(i)}{\sum_{j = 1}^n \mathbb{P}(\hat{\mathbf{H}} = \hat{\mathbf{h}} | \mathrm{X} = e_j, \mathbf{u}) \prod_{v=1}^{N^{\mathbf{u}}} f_{\mathrm{Z}_s}(\mathrm{z}^{(v)} | e_j) \mathbf{p}_{\psi}(j)}.
\end{align}

The product $\prod_{v=1}^{N^{\mathbf{u}}} f_{\mathrm{Z}_s}(\mathrm{z}^{(v)} | e_i)$ can be rewritten using the definition of $\mathrm{z}^{(v)}$:~
\begin{align}    
    &\prod_{v=1}^{N^{\mathbf{u}}} f_{\mathrm{Z}_s}(\mathrm{z}^{(v)} | e_i) = \prod_{s = 1}^S \prod_{u = 1}^{N_s^{\mathbf{u}}} f_{\mathrm{Z}_s}(\mathrm{z}_{u,s} | e_i).
\end{align}

\noindent When we explicitly write the dependencies on the channel gains, the previous formula becomes equivalent to~\eqref{eq:f_Z_usk} and can be rewritten as~
\begin{align} \label{eq:proof_P_Z}
    &\mathbb{P}(\mathbf{Z} = \mathbf{z} | \mathrm{X}_k = e_i, \hat{\mathbf{H}} = \hat{\mathbf{h}}, \mathbf{u}).
\end{align}

\noindent Combining~\eqref{eq:proof_p_psi_prime} and~\eqref{eq:proof_P_Z}, we finally obtain~
\begin{align}
    \mathbf{p}_{\psi'}(i) & = \frac{\mathbb{P}(\hat{\mathbf{H}} = \hat{\mathbf{h}} | \mathrm{X} = e_i, \mathbf{u}) \mathbb{P}(\mathbf{Z} = \mathbf{z} | \mathrm{X}_k = e_i, \hat{\mathbf{H}} = \hat{\mathbf{h}}, \mathbf{u}) \mathbf{p}_{\psi}(i)}{\sum_{j = 1}^n \mathbb{P}(\hat{\mathbf{H}} = \hat{\mathbf{h}} | \mathrm{X} = e_j, \mathbf{u}) \mathbb{P}(\mathbf{Z} = \mathbf{z} | \mathrm{X}_k = e_i, \hat{\mathbf{H}} = \hat{\mathbf{h}}, \mathbf{u}) \mathbf{p}_{\psi}(j)} \\
    & = \frac{\mathbb{P}(\mathbf{Z} = \mathbf{z}, \hat{\mathbf{H}} = \hat{\mathbf{h}} | \mathrm{X}_k = e_i, \mathbf{u}) \mathbf{p}_{\psi}(i)}{\sum_{j = 1}^n \mathbb{P}(\mathbf{Z} = \mathbf{z}, \hat{\mathbf{H}} = \hat{\mathbf{h}} | \mathrm{X}_k = e_i, \mathbf{u}) \mathbf{p}_{\psi}(j)},
\end{align}

\noindent which coincides with~\eqref{eq:p_psi_prime} and concludes the proof.

\bibliography{bibliography}{}
\bibliographystyle{IEEEtran}

\end{document}